\documentclass[11pt,fleqn]{article}       
\usepackage{geometry}
\usepackage[T1]{fontenc}
\usepackage[utf8]{inputenc}
\usepackage{authblk}
\usepackage{mathrsfs}
\usepackage[font=small,labelfont=bf]{caption}
\usepackage{graphicx}
\usepackage{multirow}
\usepackage{amsmath,amssymb,amsfonts}
\usepackage{verbatim}
\usepackage{bm}
\usepackage{color}
\usepackage{ulem}
\usepackage{mathtools}
\usepackage{cite}
\usepackage{colortbl}
\usepackage{cancel}
\definecolor{light-gray}{gray}{0.85}
\usepackage[pdftex,colorlinks,pdfpagelabels]{hyperref}
\hypersetup{
   bookmarksnumbered,
   citecolor={blue},
   linkcolor={blue},
   urlcolor={blue},
   filecolor={blue}
} 

\newcommand{\AddrTexas}{%
\textit{Department of Physics, The University of Texas at Austin, Austin, 78712 TX, USA}
}
\newcommand{\AddrStockholm}{
\textit{Oskar Klein Center for Cosmoparticle Physics, University of Stockholm, 10691 Stockholm, Sweden}
}
\newcommand{\AddrNordita}{
\textit{Nordita, KTH Royal Institute of Technology and Stockholm University, 10691 Stockholm, Sweden}
}

\usepackage{fancyhdr}
 \usepackage[bottom]{footmisc}
\fancypagestyle{plain}{%
    \fancyhead[R]{UTTG 11-2022, NORDITA 2022-056}
    
}
\date{}
\title{\Large\bf Have Pulsar Timing Arrays detected the Hot Big Bang?
 Gravitational Waves from Strong First Order Phase Transitions in the Early Universe}
\author[1,2,3]{Katherine Freese\thanks{ktfreese@utexas.edu}}
\author[1,2]{Martin Wolfgang Winkler\thanks{martin.winkler@austin.utexas.edu}}
\affil[1]{\AddrTexas}
\affil[2]{\AddrStockholm}
\affil[3]{\AddrNordita}

\begin{document}
\maketitle
\vspace*{0mm}
\begin{abstract}
The origins of matter and radiation in the universe lie in a Hot Big Bang. We present a number of well-motivated cosmologies in which the Big Bang occurs through a strong first order phase transition -- either at the end of inflation, after a period of kination (``Kination-Induced Big Bang''),  or after a second period of vacuum-domination in the early universe (``Supercooled Big Bang''); we also propose a ``Dark Big Bang'' where only the dark matter in the Universe is created in a first-order phase transition much after inflation. 
In all of these scenarios, the resulting gravitational radiation 
can explain the tentative signals reported by the NANOGrav, Parkes and European Pulsar Timing Array experiments if the reheating temperature of the Hot Big Bang, and correspondingly the energy scale of the false vacuum,  falls in the range $T_* \sim \rho_{{\rm vac}}^{1/4} $= MeV--100~GeV. 
All the same models at higher reheating temperatures will be of interest to 
upcoming ground- and space-based interferometer searches for gravitational waves at larger frequency.
\end{abstract}
\clearpage

\section{Introduction}

In standard cosmology the Hot Big Bang denotes the reheating of the universe at the end of inflation. During this process a hot plasma of particles is created containing the photons, electrons and baryons of our present universe. Guth's pioneering ``old inflation'' featured a universe trapped in a false vacuum driving the exponential expansion of space~\cite{Guth:1980zm}. The decay of the false vacuum by quantum tunnelling was meant to terminate the inflationary epoch and to transform the vacuum energy into radiation. The original inflation model, hence, already featured the idea of the Hot Big Bang occurring through a first order phase transition. Unfortunately, old inflation is plagued by the infamous ``empty universe problem''~\cite{Guth:1982pn}: sufficient inflation requires a suppressed tunneling rate. As a consequence, the phase transition would be too slow to ever complete and the universe would never enter the radiation-dominated epoch.

 The empty universe problem was resolved in slow-roll inflation~\cite{Linde:1981mu,Albrecht:1982wi} which identifies the Hot Big Bang with the perturbative or non-perturbative decay of the inflaton field, rather than with a first order phase transition.
Yet there exist equally successful theories of the early universe closer to Guth's old inflation, i.e.\ models in which inflation ends via a first order phase transition. A prime example is double field inflation~\cite{Adams:1990ds,Linde:1990gz} which features an inflaton sector comprised of two fields: one field direction requires tunneling to get from the false to the true vacuum. In the other direction the field rolls, thereby reducing the potential barrier in the tunneling direction. The tunneling rate switches from very slow to very fast, and the universe reheats suddenly and uniformly in a Big Bang phase transition. Another successful implementation of a tunneling model is chain inflation~\cite{Freese:2004vs,Freese:2005kt,Ashoorioon:2008pj}. The latter features a universe in a false vacuum similar as old inflation. However, the false vacuum decays in a series of first order phase transitions instead of just one. Each individual transition completes quickly within a fraction of a Hubble time, while all transitions together can easily support sufficient e-foldings of inflation.

But the idea of a first order Big Bang phase transition is not only tied to the inflationary epoch. In models with a unified description of inflation and dark energy~\cite{Peebles:1998qn}, the universe typically runs through a period of kination in which the universe is dominated by the kinetic energy of the quintessence field. The Hot Big Bang may then occur through a first order phase transition at the end of kination. More generally, we dub as ``Kination-Induced Big Bang'' the scenario in which an epoch of kination ends in a first order phase transition that produces the matter and radiation of our Universe.
Another complementary example consists in a strongly supercooled phase transition which is often associated with the thermal breaking of a gauge symmetry~\cite{Witten:1980ez}. Such a transition can occur long after inflation has ended and the universe was reheated. Due to the strong supercooling, the universe becomes vacuum-dominated for a second time before the phase transition converts the vacuum energy into a hot plasma. The resulting large entropy release dilutes the preexisting plasma and (virtually) all radiation we observe today stems from the supercooled transition. The latter plays the role of the Hot Big Bang in this case. Finally, we also propose the possibility that only the dark matter (and dark radiation) is created in a first order phase transition – a Dark Big Bang – while visible matter and radiation are produced earlier by the decay of the inflaton.

In this work we will present in detail these five different cosmological scenarios in which the Hot Big Bang is associated with a first order phase transition. The formation and collision of true vacuum bubbles during the phase transition induces a strong gravitational radiation signal. By determining the gravitational wave spectrum we will be able to directly link the Hot Big Bang to observational data. 

A particularly intriguing possibility is that the Big-Bang-induced gravity waves are responsible for the tentative signal reported by the NANOGrav collaboration~\cite{NANOGrav:2020bcs}. NANOGrav recently found evidence for a stochastic common-spectrum process which affects pulsar timing residuals in its 12.5-year dataset. The signal was meanwhile confirmed by the Parkes (PPTA)~\cite{Goncharov:2021oub} and the European Pulsar Timing Array (EPTA)~\cite{Chen:2021rqp,Chalumeau:2021fpz} (see also~\cite{Antoniadis:2022pcn}). While proof of the characteristic quadrupolar Hellings-Downs correlations~\cite{Hellings:1983fr} is still withstanding, these observations may amount to the first detection of a stochastic gravitational wave background. Among the most plausible sources for such a background in the sensitivity window of pulsar timing arrays are mergers of super-massive black-hole binaries~\cite{Rajagopal:1994zj,Jaffe:2002rt,Wyithe:2002ep,Sesana:2008mz,Burke-Spolaor:2018bvk,Middleton:2020asl}, a cosmic-string network in the early universe~\cite{Vilenkin:1981bx,Vachaspati:1984gt,Damour:2004kw,Siemens:2006yp,Olmez:2010bi,Ringeval:2017eww,Ellis:2020ena,Blasi:2020mfx,Buchmuller:2020lbh} and a first order phase transition~\cite{Caprini:2010xv,Schwaller:2015tja,Kobakhidze:2017mru,Nakai:2020oit,Addazi:2020zcj,Ratzinger:2020koh,Brandenburg:2021tmp,NANOGrav:2021flc,Borah:2021ocu,DiBari:2021dri,Lewicki:2021xku,Ashoorioon:2022raz} -- of which the latter is the subject of this study. By linking the phase transition properties to the Hot Big Bang, we will be able to strongly 
 constrain the parameter space. Further, we will show that a Big Bang first order phase transition can perfectly fit the pulsar timing signals.
Fig.~\ref{fig:spectra2} shows our main results of matching predictions of our five cosmological models to the data.  Needless to say that a direct experimental probe of the Hot Big Bang would be of paramount importance.

The paper is organized as follows: in Sec.~\ref{sec:gravitywaves} we review the calculation of the gravitational wave spectrum from a first order phase transition. The derivation of the time, duration and strength of the phase transition (entering the spectrum) are also provided. In Sec.~\ref{sec:nanograv} we perform a fit to the pulsar timing signal with the focus on a Big Bang phase transition. In Sec.~\ref{sec:scenarios} we describe several cosmological scenarios in which the Big Bang occurs through a first order phase transition. We also determine the corresponding gravitational wave signals and show that they can potentially explain the pulsar timing data. Finally, Sec.~\ref{sec:conclusion} contains our concluding remarks.

\section{Gravity Waves from a First Order Phase Transition}\label{sec:gravitywaves}

\subsection{Gravitational Wave Spectrum}

We consider a first order phase transition in the early universe triggered by the decay of a false vacuum with energy density $\rho_{\text{vac}}$. Following the standard convention, we introduce the parameter~\cite{Kamionkowski:1993fg}
\begin{equation}\label{eq:alpha}
\alpha = \frac{\rho_{\text{vac}}}{\rho_{\text{r}}(T_n)}\,,
\end{equation}
which specifies the ratio of the vacuum energy density to the energy density of the surrounding radiation plasma characterized by its temperature $T_n$ right before the transition,
\begin{equation}
\rho_{\text{r}}(T_n) = \frac{\pi^2}{30}g_{\text{eff}}(T_n)T_n^4\,,
\end{equation}
where $g_{\text{eff}}$ denotes the effective number of relativistic species. The special case of a phase transition in vacuum (i.e.\ without any preexisting plasma) corresponds to $T_n=0$ and $\alpha\rightarrow \infty$.

During the phase transition, bubbles of true vacuum are formed at random nucleation sites which quickly grow and collide with other bubbles. In this process the universe is reheated, i.e. the vacuum energy is converted to thermal energy of the radiation plasma. We denote the temperature of the radiation bath right after the transition by $T_*$. If the transition time is short (compared to the Hubble time) we can approximate,
\begin{equation}\label{eq:rhotot}
\rho_{\text{r}}(T_*) \simeq \rho_{\text{tot}} \simeq \rho_{\text{r}}(T_n) + \rho_{\text{vac}},
\end{equation}
where $\rho_{\text{tot}}$ stands for the total energy density at the phase transition. This implies,
\begin{equation}\label{eq:Tst}
 T_* \simeq\left(\frac{30}{\pi^2 g_{\text{eff}}(T_*)}\left(\rho_{\text{vac}}+\rho_{\text{r}}(T_n)\right)\right)^{1/4}=\left(\frac{\alpha+1}{\alpha}\right)^{1/4}\left(\frac{30\,\rho_{\text{vac}}}{\pi^2 g_{\text{eff}}(T_*)}\right)^{1/4}\,.
\end{equation}
In the case of a phase transition in vacuum $\alpha\rightarrow\infty$ and the factor $(\alpha+1)/\alpha$ simply becomes unity.

First order phase transitions can source strong gravitational radiation~\cite{Witten:1984rs,Hogan:1986qda} which is generated by the collisions of true vacuum bubbles~\cite{Kosowsky:1992rz,Kosowsky:1992vn} as well as sound waves~\cite{Hindmarsh:2013xza,Hindmarsh:2015qta,Hindmarsh:2017gnf} and magneto-hydrodynamic turbulence in the surrounding plasma induced by the expanding bubbles~\cite{Kosowsky:2001xp,Dolgov:2002ra,Caprini:2009yp}. The relative importance of the different contributions depends on the underlying microphysics. In the following we will mostly focus on the case $\alpha\gtrsim 1$, in which the vacuum decay generates most (or all) of the radiation plasma in the universe, while the preexisting plasma is subdominant (or absent). Assuming, furthermore, that the field undergoing the phase transition does not couple strongly to the radiation plasma (if present), we expect the bubbles to propagate at the speed of light and their collisions to be the dominant source of gravitational radiation~\cite{Espinosa:2010hh}. A possible exception occurs if the phase transition is connected with the breaking of a gauge symmetry. The radiation of soft gauge bosons inflicts a pressure on the bubble walls which grows linearly with their Lorentz boost~\cite{Bodeker:2017cim} (or even quadratically~\cite{Hoche:2020ysm}). In this case the bubble walls may lose most of their energy to the surrounding plasma even if $\alpha \gg 1$ such that the gravitational wave emission is dominated by plasma processes.

The gravitational wave spectrum today, induced by bubble collisions at a phase transition in the early universe, normalized to the critical density today as a function of frequency $f$ takes the form~\cite{Kosowsky:1992rz,Kosowsky:1992vn}
\begin{equation}\label{eq:gravityspectrum}
\Omega_{\text{GW}} h^2(f)  = \left(\frac{7.6\times 10^{-5}}{g^{1/3}_{\text{eff}}(T_*)}\right)\;\,\widetilde{\Omega}\;\left(\frac{H_*}{\beta}\right)^2\,\left(\frac{\kappa_\phi \alpha}{1+\alpha}\right)^2\, \frac{(a+b)\left(f/f_{\text{peak}}^0\right)^a}{b+a\left(f/f_{\text{peak}}^0\right)^{a+b}}
\,,
\end{equation}
where we set the bubble wall velocity to the speed of light (which is valid for all scenarios discussed in this work). 
The expected spectrum corresponds to a (smoothly) broken power law with a maximum at the redshifted peak frequency $f_{\text{peak}}^0$. The parameter $\widetilde{\Omega}$ sets the overall normalization of the spectrum, while $a$, $b$ determine the power law index in the infrared ($f<f_{\text{peak}}^0$) and ultraviolet ($f>f_{\text{peak}}^0$) respectively.
The expected values of these quantities from simulations of bubble collisions (shown in Tab.~\ref{tab:gwparameters}) will be discussed in more detail shortly.
The first term in brackets on the right-hand side of Eq.(\ref{eq:gravityspectrum}) accounts for the redshift of the gravity wave amplitude from production until now. 
We note that the gravitational wave amplitude depends on the square of the factor 
\begin{equation}
\alpha / ( 1 + \alpha) = \rho_{\text{vac}} / \rho_{\text{tot}}\, .
\end{equation}  
Furthermore, $H_*$ is the Hubble rate at the phase transition, while $\beta$ stands for the inverse time duration of the phase transition (a precise definition of $\beta$ will follow in Eq.~\eqref{eq:beta}). 
The gravitational wave amplitude depends on the quantity $(H_*/\beta)$, the number of e-foldings (of the scale factor) during the phase transition.  In this paper, as we will see, we will always be driven to $(H_*/\beta) <1$, a requirement that suppresses the gravitational wave amplitude.

The quantity $\beta$ also determines the peak frequency of the gravitational wave spectrum at the time of production~\cite{Huber:2008hg},
\begin{equation}\label{eq:peakf_emission}
 f_{\text{peak}} \simeq 0.2\, \beta\,,
\end{equation}
which in the present universe has redshifted to the value,
\begin{equation}\label{eq:peakf}
f_{\text{peak}}^0 \simeq 7.7\times 10^{-8}\:\text{Hz}\;\,\left(\frac{f_{\text{peak}}}{H_*}\right)\,\left(\frac{g^{1/6}_{\text{eff}}(T_*)\,T_*}{\text{GeV}}\right)\,.
\end{equation}
The parameter $\kappa_\phi$ in~Eq.~\eqref{eq:gravityspectrum} specifies the energy fraction carried by the bubble walls at collision. For phase transitions in vacuum or with negligible impact of the surrounding plasma one can simply set $\kappa_\phi=1$. 

A particular relevant special case is a first order phase transition in vacuum (e.g.\ at the end of inflation). In the absence of a preexisting plasma the factor $\alpha / (1+ \alpha) =\rho_{\text{vac}}/\rho_{\text{tot}}$ in the gravitational wave spectrum simply becomes unity (cf.~Eq.~\eqref{eq:alpha} and~\eqref{eq:rhotot}). In order to find a rough estimate for the gravity wave amplitude at the peak frequency in the pure vacuum case, we approximate $g_{\text{eff}}(T_*)=10$ and $\widetilde{\Omega}= 0.05$ to find
\begin{equation}\label{eq:vacuumomega}
\Omega_{\text{GW}} h^2(f_{\text{peak}}^0)  \sim 1.8\times 10^{-6}\;\left(\frac{H_*}{\beta}\right)^2\quad \text{(vacuum phase transition)}\,.
\end{equation}
Note that the gravitational wave amplitude in this case is completely determined by the quantity $(H_*/\beta)$, the number of e-foldings (of the scale factor) during the tunneling transition. As mentioned above, in this paper we will find the requirement $(H_*/\beta) <1$, leading to suppression of the gravitational wave amplitude. The peak frequency for the pure vacuum case can be estimated as,
\begin{equation}\label{eq:vacuumf}
f_{\text{peak}}^0 \sim 1.7\times 10^{-8}\:\text{Hz}\;\,\left(\frac{\beta}{H_*}\right)\,\left(\frac{\rho_{\text{vac}}^{1/4}}{\text{GeV}}\right)\quad \text{(vacuum phase transition)}\,.
\end{equation}
From Eq.(\ref{eq:gravityspectrum}), we can see that the largest value of the gravitational wave amplitude $\Omega_{\text{GW}}$ is achieved for the pure vacuum case, in which $\alpha \rightarrow \infty$ so that the factor $\alpha / ( 1 + \alpha)$ takes its largest possible value of unity.  Below (see Eq.~\eqref{eq:percolation}) we will require $H_*/\beta < 1/3$; with this requirement,  Eq.(\ref{eq:gravityspectrum}) leads to a maximum predicted value $\Omega_{\text{GW}} < 10^{-7}$. 
Previously~\cite{Schmitz:2020syl} studied a variety of benchmark cases in agreement with this upper bound.

Let us now also briefly turn to the second potential source of gravitational radiation, which are sound waves in the plasma induced by the expanding vacuum bubbles. The corresponding acoustic gravitational wave spectrum has been computed to be~\cite{Hindmarsh:2015qta,Caprini:2018mtu},
\begin{equation}\label{eq:gravityspectrumplasma}
\Omega_{\text{GW}} h^2(f)  = \left(\frac{7.6\times 10^{-5}}{g^{1/3}_{\text{eff}}(T_*)}\right)\;\left(\frac{H_*}{\beta}\right)\,\left(\frac{\kappa_v \alpha}{1+\alpha}\right)^2\, 
\left(\frac{f}{f_{\text{peak}}^0}\right)^a\left(\frac{7}{4+3(f/f_{\text{peak}}^0)^2}\right)^{\frac{b+a}{2}}
\,,
\end{equation}
where $\kappa_v$ denotes the fraction of vacuum energy which is converted into bulk motion of the plasma. According to the recent simulation~\cite{Hindmarsh:2017gnf}, the peak frequency of the gravitational waves from sound waves is very similar to the one from bubble collisions\footnote{A somewhat higher peak frequency $f_{\text{peak}} \simeq 1.15 \beta$ of the acoustic gravitational wave spectrum had previously been suggested in~\cite{Hindmarsh:2015qta}}. Therefore,~Eq.~\eqref{eq:peakf_emission} and~\eqref{eq:peakf} can also be applied for the acoustic gravitational wave spectrum in Eq.~\eqref{eq:gravityspectrumplasma}. 

Vacuum bubbles expanding through a plasma can also induce magneto-hydrodynamic turbulence which is another possible source of gravitational waves~\cite{Kosowsky:2001xp,Dolgov:2002ra,Caprini:2009yp}. Since this contribution suffers from a high degree of uncertainty we will not explicitly consider it in this work (but we will comment in case of relevance).

Let us now discuss in more detail the frequency-dependence of the gravitational spectrum from bubble collisions in Eq.~\eqref{eq:gravityspectrum} and from sound waves in Eq.~\eqref{eq:gravityspectrumplasma}. 
In both cases the expected spectrum peaks at the redshifted peak frequency $f_{\text{peak}}^0$ with
 $a$, $b$ giving the power law indices below and above the peak respectively. As above, in both cases the parameter $\widetilde{\Omega}$ sets the overall normalization of the spectrum.
In Tab.~\ref{tab:gwparameters} we provide the parameters obtained via numerical simulation of bubble collisions and sound waves. In the case of bubble collisions we separately quote the result of the envelope approximation~\cite{Kosowsky:1992rz,Kosowsky:1992vn,Huber:2008hg} and of the lattice simulation~\cite{Cutting:2020nla} which we denote as `thick-wall simulation' in the following.

\begin{table}[t]
\begin{center}
\begin{tabular}{|cccc|}
\hline
&&&\\[-4mm]
 & $\quad\widetilde{\Omega}\quad$ & $\quad a\quad$ & $\quad b \quad$ \\
 \hline
&&&\\[-4mm] 
 envelope & $0.077$ & $2.8$ &  $1$ \\
 thick-wall & $0.027$ & $0.7$ &  $2.2$\\ 
 sound waves & $0.16$ & $3$ &  $4$\\ \hline
\end{tabular}
\end{center}
\vspace{-0.4cm}
\caption{Parameters entering the gravitational wave spectrum from bubble collisions in a first order phase transition (Eq.~\ref{eq:gravityspectrum}) as determined in the envelope approximation (taken from~\cite{Huber:2008hg}) and in the thick-wall simulation~\cite{Cutting:2020nla}. Also quoted are the parameters entering the acoustic gravitational wave spectrum given in Eq.~\eqref{eq:gravityspectrumplasma}~\cite{Hindmarsh:2015qta}.}
\label{tab:gwparameters}
\end{table}

In the envelope approximation, the stress-energy is assumed to be located in a thin shell at the bubble wall which disappears upon collision. The gravitational radiation is sourced only by the uncollided envelope of the spherical bubbles, ignoring the interaction region. The envelope approximation is expected to apply to phase transitions in which the tunneling field becomes trapped temporarily in the false vacuum within the bubble collision region (which justifies the neglect of the shear stress after collision)~\cite{Jinno:2019bxw}. This has been shown to occur in the thin-wall regime of vacuum tunneling, i.e.\ when the energy difference between the false and the true vacuum is small compared to the potential barrier separating the two~\cite{Hawking:1982ga,Watkins:1991zt,Falkowski:2012fb}. However, in the opposite thick-wall regime, the tunneling field does not get trapped and rather undergoes oscillations around the true vacuum in the bubble overlap region. This leads to significant propagation of the shear stress after collision -- strongly violating the basic assumptions of the envelope approximation~\cite{Cutting:2020nla}. In the thick-wall case the gravitational wave spectrum was argued~\cite{Jinno:2019bxw} to follow more closely the predictions of the bulk flow model~\cite{Konstandin:2017sat} in which the shell of shear-stress continues to propagate after collision. This picture was qualitatively confirmed by a recent lattice simulation which included an explicit modeling of the field profile during the bubble collision stage assuming a quartic potential~\cite{Cutting:2020nla}. The parameters obtained there for the thick-wall case\footnote{The thick-wall case corresponds to the smallest $\bar{\lambda}$ simulated in~\cite{Cutting:2020nla}.} shown in Tab.~\ref{tab:gwparameters} are in reasonable agreement with the predictions of the bulk flow model.

A striking observation is that the gravity wave spectrum rises more steeply in the infrared and falls more softly in the ultraviolet region in the envelope approximation compared to the thick-wall simulation. This difference is not unexpected since both derivations describe different physical realities (thin-wall bubbles vs. thick-wall bubbles). Note, however, that in both cases the simulations were optimized to predict the gravity wave spectrum around the peak frequency and may not capture well the behavior in the far-infrared ($f\ll f_{\text{peak}}^0$) and far-ultraviolet ($f\gg f_{\text{peak}}^0$) regime. Causality considerations suggest a power law index $a\rightarrow 3$ for $f\ll H_*$ (see e.g.~\cite{Caprini:2009fx}) hinting at a transition to a steeper power law at very low frequency not resolved in the simulations.

\subsection{Phase Transition Parameters}

The time and the duration of a first order phase transition can be linked to to the false vacuum decay rate per volume $\Gamma$. In the microphysical realization, the latter corresponds to the transition rate of a scalar field between two minima of its potential. One finds~\cite{Coleman:1977py,Callan:1977pt,Linde:1980tt,Linde:1981zj}
\begin{equation}\label{eq:tunnelingrate}
 \Gamma \simeq \text{max}\left[m^4 \left(\frac{S_4}{2\pi}\right)^2 e^{-S_4},\:
 T^4 \left(\frac{S_3}{2\pi\,T}\right)^{3/2} e^{-S_3/T} \right]\,,
\end{equation}
where $S_4$ and $S_3$ stand for the 4- and 3-dimensional Euclidean actions of the bounce solution extrapolating between the two vacua, while $m$ is the mass of the scalar field (evaluated in the false vacuum).

The first term in~Eq.~\eqref{eq:tunnelingrate} corresponds to the quantum tunneling rate at zero temperature, while the second term is the thermally induced rate. In the absence of a preexisting plasma (i.e. if the phase transition occurs in vacuum), $\Gamma$ is given by the quantum tunneling rate. If a plasma with temperature $T$ is present, $\Gamma$ is determined by the faster of the two rates.

The probability $P(t)$ of finding a point in the false vacuum at the time $t$ can be determined by integrating $\Gamma$ over the past light cone of the point~\cite{Guth:1979bh,Guth:1981uk},
\begin{equation}\label{eq:prob}
P(t) = e^{-I(t)}\,,\qquad I(t)=\frac{4\pi}{3}\int\limits_0^t dt^\prime\, \Gamma(t^\prime) a^3(t^\prime) r_{\rm com}^3(t,t^\prime)\,,
\end{equation}
where $I(t)$ corresponds to the expected number of bubble nucleation sites in the past light cone. The time of the phase transition $t_*$ can be defined as the (mean) decay time of the false vacuum,
\begin{equation}\label{eq:Ieq1}
I(t_*)=1\,.
\end{equation}
In Eq.~\eqref{eq:prob} the comoving radius of the past light cone $r_{\rm com}$ is obtained as,
\begin{equation}
\label{eq:comovingradius}
r_{\rm com}(t,t^\prime)= \int\limits_{t^\prime}^t \frac{d\tilde{t}}{a(\tilde{t})}\,,
\end{equation}
where the scale factor $a(t)$ of a universe containing vacuum energy and radiation reads,
\begin{equation}
a(t)=a(t_0) \exp\left(\int\limits_{t_0}^t dt^\prime H(t) dt^\prime\right)\,,\qquad H(t)=\sqrt{\frac{\rho_{\text{vac}}+\rho_{\text{r}}(t)}{3\,M_{\text{P}}^2}}\,.
\end{equation}

The duration of the phase transition $\beta^{-1}$ depends on how quickly the false vacuum probability $P(t)$ decreases with time. A convenient definition is,
\begin{equation}\label{eq:beta}
\beta=-\left.\frac{\dot{P}}{P}\right|_{t=t_*} =\left.\dot{I}\,\right|_{t=t_*}\,.
\end{equation}
We can compute $\beta$ for the two  cases of quantum tunneling at finite temperature and at zero temperature.
If the false vacuum decay rate $\Gamma$ in~Eq.~\eqref{eq:tunnelingrate} is set by the thermal transition rate, it exhibits a strong exponential time-dependence (through the temperature of the plasma). In this case the time dependence of $I(t)$ in Eq.~\eqref{eq:prob} is determined primarily by the exponential time-dependence of $\Gamma$ (rather than by the power-law time-dependence of the light cone volume).  Hence $\beta \sim \dot{\Gamma} / \Gamma\big|_{t=t_*} $.
 In contrast, if the field $\phi$ driving the phase transition is (almost) decoupled from the surrounding plasma or if the phase transition occurs in vacuum, $\Gamma$ is set by the zero-temperature quantum tunneling rate. The latter is time-independent in the simplest case, where tunneling is not affected by other interactions of the tunneling field. For cases of (nearly) constant $\Gamma$, the change of the four-volume of the past light cone in~Eq.~\eqref{eq:prob} determines $\beta$. Note, however, that vacuum tunneling does not generically imply $\Gamma=\text{const}$. This is because a strong exponential time-dependence of $\Gamma$ can also arise if the tunneling field couples to a spectator field with a time-dependent evolution. Hence, for vacuum tunneling, it depends on the underlying model whether the upper or lower expression in~Eq.~\eqref{eq:beta2} below applies.
In summary,
\begin{equation}\label{eq:beta2}
\beta\simeq 
\begin{cases} \left.\frac{\dot{\Gamma}}{\Gamma}\right|_{t=t_*} & \Gamma\neq \text{const}\,,\\[3mm]
\frac{4\pi\Gamma}{a(t_*)}\int\limits_0^{t_*} dt^\prime\,  a^3(t^\prime)\, r_{\rm com}^2(t_*,t^\prime)  
& \Gamma\simeq \text{const} \,.\end{cases}
\end{equation}

The successful completion of a first order phase transition requires the bubbles of true vacuum to percolate such that the energy of the bubble walls can be transferred into radiation. It may naively seem that $\beta>0$ -- i.e.\ a decreasing probability of a point to stay in the false vacuum -- would automatically ensure percolation. However, this is not true since the physical volume of the false vacuum $V_{\text{false}}\propto a^3(t) P(t)$ may increase even for decreasing $P(t)$ due to $a(t)$ growing by the Hubble expansion~\cite{Ellis:2018mja}. Therefore, the relevant criterion for effective bubble percolation is that $V_{\text{false}}$ decreases around the time of the phase transition $t_*$~\cite{Turner:1992tz},
\begin{equation}\label{eq:percolation}
\left.\frac{d}{dt} (a^3 P)\right|_{t=t_*} < 0\quad\Longrightarrow\quad \beta > 3H_*\,.
\end{equation}
The above condition limits the amplitude of gravitational wave emission by a first order phase transition which scales with $(H_*/\beta)^2$.

\begin{figure}[t!]
\begin{center}
\includegraphics[width=0.5\textwidth]{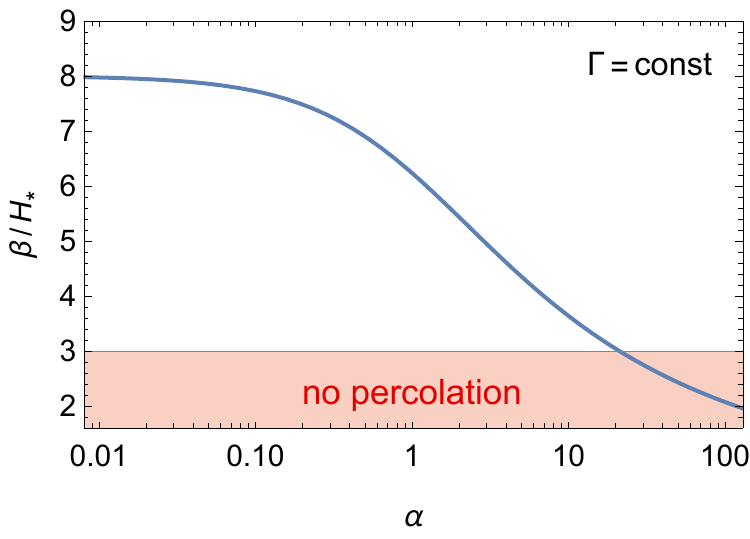}
\end{center}
\vspace{-5mm}
\caption{Inverse duration $\beta$ of a first order phase transition for a time-independent vacuum decay rate. Here $\alpha$ is the ratio of the vacuum energy density to the energy density of the surrounding radiation plasma right before the transition. In the red region the bubble percolation condition is violated.}
\label{fig:betaalpha}
\end{figure}

In the special case $\Gamma=\text{const}$ the duration of the phase transition can be calculated explicitly from Eq.~\eqref{eq:beta2}. The resulting $\beta$ as a function of $\alpha$ is shown in Fig.~\ref{fig:betaalpha}. As a reminder, $\alpha$ is the ratio of the vacuum energy density to the energy density of the surrounding radiation plasma right before the transition (see Eq.~\eqref{eq:alpha});  for a single first order phase transition as in ``old inflation'', $\alpha \rightarrow \infty$. It can be seen that the bubble percolation condition imposes an upper limit $\alpha \lesssim 20$ by which vacuum energy dominates over the preexisting plasma in a successful phase transition with constant $\Gamma$.\footnote{A similar conclusion for cases with a slowly varying $\Gamma$ was drawn in~\cite{Ellis:2018mja}.} Note, however, that this constraint does not apply to cases with $\Gamma\neq \text{const}$ for which $\alpha$ can take any value (including $\alpha=\infty$ as for a phase transition in vacuum).

\section{Pulsar Timing Array Signal from a Phase Transition}\label{sec:nanograv}

The NANOGrav, PPTA and EPTA collaborations have reported strong evidence for a spectrally-similar low-frequency stochastic process which affects pulsar timing residuals~\cite{NANOGrav:2020bcs,Goncharov:2021oub,Chen:2021rqp}. Searches for the quadrupolar Hellings-Downs correlations~\cite{Hellings:1983fr} which would establish a gravitational wave origin are not yet conclusive due to limited statistics. However, the spectral properties of the signal are consistent with a stochastic gravitational wave background at frequencies $f\sim 1\:\text{yr}^{-1}$. 

\subsection{Fitting the Pulsar Timing Signal}\label{sec:fitting}
\begin{figure}[h]
\begin{center}
\includegraphics[width=0.6\textwidth]{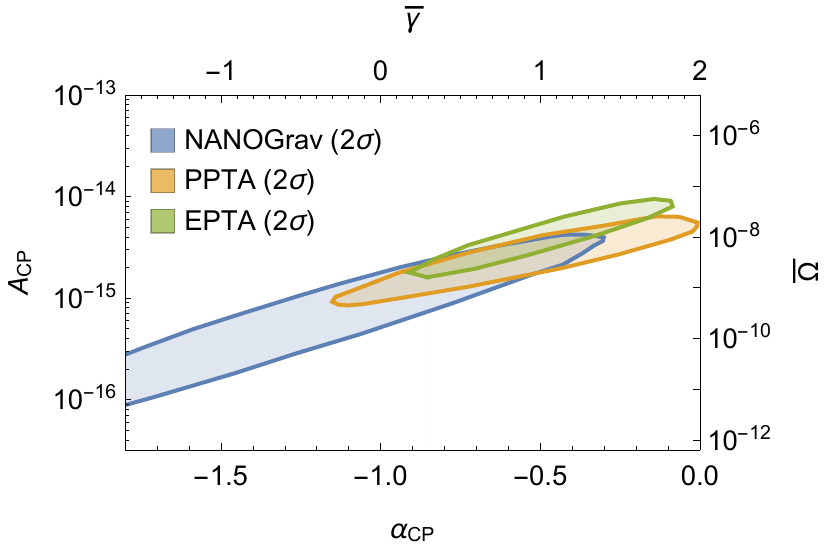}
\end{center}
\vspace{-5mm}
\caption{The signals of NANOGrav, PPTA and EPTA interpreted as a stochastic gravitational wave background. Shown are the 2$\sigma$-preferred regions for the power spectrum of the characteristic strain modelled as a power law (cf. Eq.~\eqref{eq:charstrain}). The preferred region can directly be mapped to the gravity wave spectrum in terms of the critical density. Corresponding parameters (as defined in Eq.~\eqref{eq:omegapl}) are also shown on the axes.}
\label{fig:nanogravPL}
\end{figure}

Within the accessible frequency band, the observed power spectrum of the characteristic strain $h_c(f)$ is consistent with a power law,
\begin{equation}\label{eq:charstrain}
 h_c(f) = A_{\text{CP}} \:\left(\frac{f}{\text{yr}^{-1}}\right)^{\alpha_{\text{CP}}}\,.
\end{equation}
The preferred regions in terms of the power law index $\alpha_{\text{CP}}$ and the normalization $A_{\text{CP}}$ obtained in the NANOGrav, PPTA and EPTA analyses~\cite{NANOGrav:2020bcs,Goncharov:2021oub,Chen:2021rqp} are shown in Fig.~\ref{fig:nanogravPL}.

The power spectrum of the characteristic strain is directly related to the gravity wave spectrum in terms of the critical density,
\begin{equation}\label{eq:omegahc}
 \Omega_{\text{GW}} h^2(f) = \frac{2\pi^2}{3 (H_0/h)^2} f^2  \,h_c^2(f)\,,
\end{equation}
where $H_0 = 100\,h\:\text{km}\,\text{s}^{-1}\text{Mpc}^{-1}$ denotes the Hubble constant. If we model $\Omega_{\text{GW}} h^2(f)$ as a power law,
\begin{equation}\label{eq:omegapl}
 \Omega_{\text{GW}} h^2(f) = \overline{\Omega} \left(\frac{f}{\text{yr}^{-1}}\right)^{\bar{\gamma}}\,,
\end{equation}
Eq.~\eqref{eq:omegahc} allows us to directly map the preferred region from the $\alpha_{\text{CP}}$-$A_{\text{CP}}$-plane into the $\bar{\gamma}$-$\overline{\Omega}$ plane as is also shown in Fig.~\ref{fig:nanogravPL}.

We now turn to the interpretation of the pulsar timing signals in terms of a first order phase transition. The corresponding gravitational wave spectrum follows a broken power law with the break (= a maximum) at the peak frequency $f_{\text{peak}}^0$ (see Eq.~\eqref{eq:gravityspectrum}). In most of the parameter space $f_{\text{peak}}^0$ falls outside the frequency band of the pulsar timing arrays which would only measure the rising or falling part of the spectrum and, hence, a single power law. This means that the analyses for the power law case as shown in Fig.~\ref{fig:nanogravPL} can directly be applied. In order to cover also cases with the peak of the gravity wave spectrum inside the experimental frequency bands, we use the following procedure to derive an `average power law': 
\begin{enumerate}
\item we determine the power law index $\gamma_i$ and the normalization parameter separately for each of the measurement frequencies
\begin{equation}
\gamma_i=\left.\frac{d\log \Omega_{\text{GW}}}{ d\log f}\right|_{f=f_i}\,\qquad
\Omega_i=\left.\frac{\Omega_{\text{GW}} h^2}{ (f/\text{yr}^{-1})^{\gamma_i}}\right|_{f=f_i}
\end{equation}
\item we define the averaged power law index and normalization by weighting the $\gamma_i$ and $\Omega_i$ with the experimental sensitivity $w_i$ at each of the frequencies,
\begin{equation}
\bar{\gamma} = \sum\limits_{i=1}^5 w_i \gamma_i\,\qquad \log\overline{\Omega} = \sum\limits_{i=1}^5 w_i \log\Omega_i\,.
\end{equation}
The $w_i$ are approximated by the inverse error in the frequency bin normalized such that $\sum_i w_i =1$.
\end{enumerate}
Defining an averaged amplitude and power law index is only reasonable for a small number of measurement frequencies in a relatively narrow band. Therefore, we will only apply the described method to the NANOGrav and PPTA 5-frequency data sets and not to EPTA, for which only a 30-frequency analysis is available.\footnote{The 5-frequency PPTA analysis is presented in Fig.\ 1 (left panel) of~\cite{Goncharov:2021oub}. The error in each bin (which determines $w_i$) is extracted from the right panel of the same figure for PPTA and from the interactive version of Fig. 2 in~\cite{NANOGrav:2021flc} for NANOGrav.}

After matching the gravitational wave spectrum to the effective power law form we can apply the constraints from~\cite{NANOGrav:2020bcs} as shown in Fig.~\ref{fig:nanogravPL}. This allows us to estimate the NANOGrav and PPTA signal regions for a first order phase transition (see Fig.~\ref{fig:spectra}). The signal region for EPTA, which we do not explicitly derive (for the reason stated above), is expected to fall in a very similar range.

\subsection{Implications for a first order Big Bang Phase Transition}\label{sec:implications}

Our main focus is on cosmological scenarios in which the Big Bang occurs through a first order phase transition. This goes back to Guth's seminal idea that the universe was initially trapped in a metastable vacuum driving cosmic inflation~\cite{Guth:1980zm}. Quantum tunneling into the true vacuum then triggers a first order phase transition which was meant to terminate the inflationary epoch. Guth's original old inflation model, however, suffers from the empty-universe problem -- the phase transition is too slow to ever complete~\cite{Guth:1982pn}. True vacuum bubbles are formed so distantly that they never percolate and reheat the universe. 

Yet, the failure of old inflation does not rule out a first order Big Bang phase transition which produces all (or most of) the matter and radiation in our present universe. Old inflation corresponds to a phase transition with $\Gamma=\text{const}$, $\alpha=\infty$. Relaxing any of these two assumptions -- i.e. considering a time-dependent tunneling rate and/or a (subdominant) preexisting radiation plasma -- can reconcile a Big Bang phase transition with the percolation condition. We will later present a number of well-motived cosmological scenarios with these properties. In order to capture a wide class of Big Bang phase transitions, we will thus consider the following two cases:
\begin{enumerate}
\item A phase transition in vacuum ($\alpha=\infty$) with a time-dependent vacuum decay rate $\Gamma\neq\text{const}$.
\item a phase transition within a preexisting plasma ($\alpha\neq \infty$) with a constant vacuum decay rate $\Gamma=\text{const}$.
\end{enumerate}

\begin{figure}[t]
\begin{center}
\includegraphics[width=0.48\textwidth]{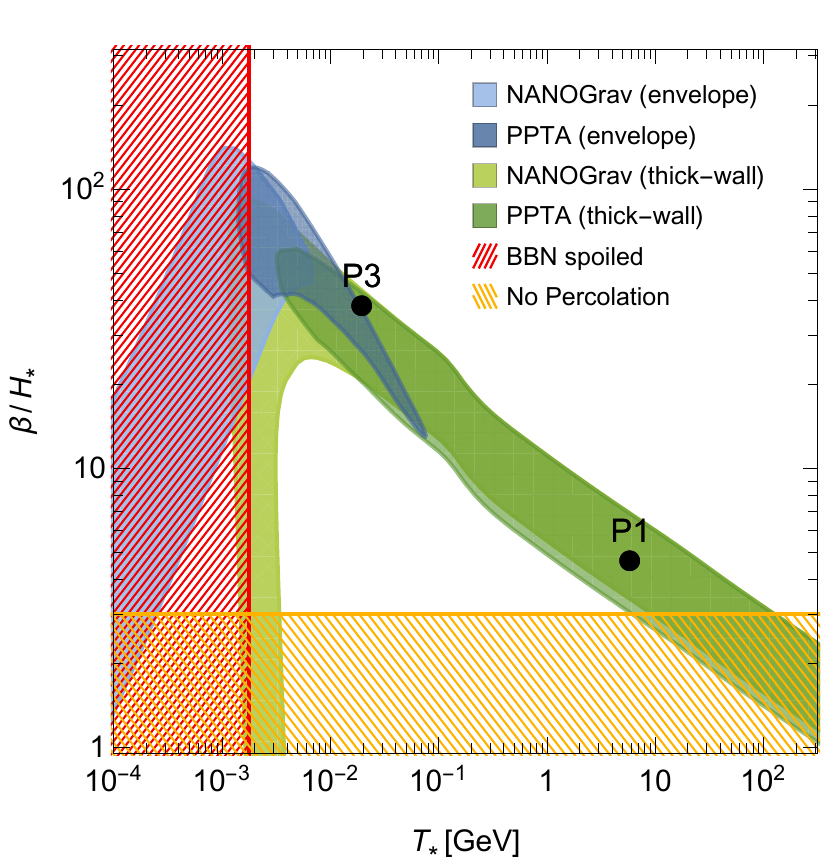}
\includegraphics[width=0.48\textwidth]{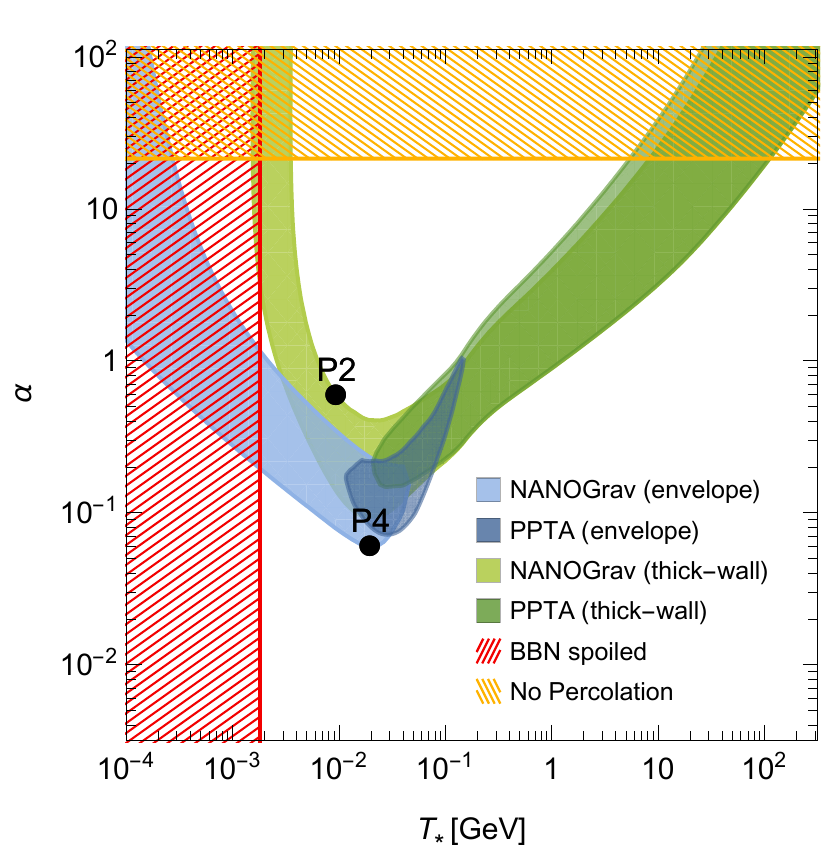}
\end{center}
\caption{Parameter regions in which a first order phase transition can explain the tentative gravitational wave signal observed by NANOGrav (light colored) and PPTA (dark colored). Preferred regions are depicted separately for the gravitational wave spectrum of the envelope approximation~\cite{Kosowsky:1992rz,Huber:2008hg} (blue) and the thick-wall simulation~\cite{Cutting:2020nla} (green). Observational constraints from BBN and bubble percolation are also shown. The left panel corresponds to a phase transition in vacuum ($\alpha=\infty$). The right panel corresponds to a phase transition with constant vacuum decay rate $\Gamma=\text{const}$ (for this case $\beta$ is fixed by $\alpha$ as shown in Fig.~\ref{fig:betaalpha}). The benchmark points P1-P4 are as defined in Fig.~\ref{fig:spectra2}, which shows the gravitational wave spectrum for these points.}

\label{fig:spectra}
\end{figure}

For Big Bang phase transitions we can focus on the gravitational wave spectrum from bubble collisions given in Eq.~\eqref{eq:gravityspectrum} (with $\kappa_\phi$ set to unity). In Fig.~\ref{fig:spectra} we present the range of ($T_*$, $\beta$) and ($T_*$, $\alpha$) for which the NANOGrav and PPTA signals can be explained by a Big Bang phase transition for the two cases described above. The signal regions (following from the derivation in Sec.~\ref{sec:fitting}) are depicted separately for the gravity wave emission predicted by the envelope approximation and by the thick-wall simulation (cf. Tab.~\ref{tab:gwparameters}). Also shown in the figure are the constraints imposed by bubble percolation (Eq.~\eqref{eq:percolation}) and by primordial nucleosynthesis (BBN). Successful BBN requires the phase transition to reheat the universe to a temperature $T_*>1.8\:\text{MeV}$~\cite{Hannestad:2004px,Hasegawa:2019jsa}.

A striking observation is that the favored phase transition temperature $T_*$ (or correspondingly the energy scale of the phase transition) strongly depends on the implemented gravitational wave spectrum. If the spectrum follows the envelope approximation, $T_*\lesssim 100\:\text{MeV}$ is required to fit the pulsar timing signal -- just barely consistent with BBN. In contrast a higher $T_*\simeq \text{MeV}-100\:\text{GeV}$ is preferred for the spectrum predicted by the thick-wall simulation. The origin of this discrepancy is easy to understand: the envelope approximation predicts the gravitational wave spectrum to rise with a power law index $a=2.8$ for $f<f_{\text{peak}}^0$ which is outside the NANOGrav, PPTA and EPTA $2\sigma$-windows independent of the amplitude (see Fig~\ref{fig:nanogravPL}). Therefore, in order to fit the pulsar timing signal in the envelope approximation, the peak frequency $f_{\text{peak}}^0$ must reside inside or below the covered frequency band ($f\simeq 1-10\:\text{nHz}$). This translates to the upper limit of $T_*$ in the MeV-range (cf.\ Eq.~\eqref{eq:peakf}). 

The thick-wall simulation, on the other hand, predicts a much softer gravity wave spectrum in the infrared (power law index $a=0.7$) consistent with the pulsar timing signal. At the same time, the spectrum falls very quickly in the ultraviolet (power law index $b=2.2$) which strongly suppresses the signal above $f_{\text{peak}}^0$. Hence -- contrary to the envelope approximation -- the thick-wall simulation favors a peak frequency within or above the frequency band of the pulsar timing arrays. This general trend has already been noted in a previous analysis~\cite{NANOGrav:2021flc}.\footnote{The consistency of NANOGrav with $T_*> \text{GeV}$ is noted in the main text of~\cite{NANOGrav:2021flc}, but due to the specific priors not fully visible in Fig.\ 1 of this reference, which shows the preferred NANOGrav region for gravity waves from bubble collisions (in blue).} 

Compared to NANOGrav, PPTA and EPTA have a stronger preference for a rising ($\bar{\gamma}\gtrsim 0$) gravitational wave spectrum at the measured frequencies (see Fig~\ref{fig:nanogravPL}). Therefore, some of the parameter space at low $T_*$ consistent with NANOGrav does not provide a good fit for the other two pulsar timing arrays (for PPTA this is directly visible in Fig.~\ref{fig:spectra}). Since most of the low-$T_*$-regime is, however, anyway excluded by BBN this difference is of minor importance.

\subsection{Properties of Potentials for First Order Transitions Required by Pulsar Timing Array Data}\label{sec:properties}

We can now ask what the implications of our main results illustrated in Fig.~\ref{fig:spectra} are for general properties of potentials that are responsible for first order phase transitions.  Five examples of potentials will be discussed in the subsequent sections, but 
we can already determine what the scales of the potentials must be in order to explain the pulsar timing array data. 

For this purpose we use~Eq.~\eqref{eq:Tst} in order to obtain the preferred range of $\rho_{\text{vac}}$ for the signal regions shown in Fig.~\ref{fig:spectra}. For the gravitational wave spectrum from the envelope approximation and the thick-wall simulation we find
\begin{equation}
\rho_{\text{vac}}^{1/4}\simeq \begin{cases}
2\:\text{MeV} - 0.2 \:\text{GeV}& \text{(envelope approximation)\,,}\\
2\:\text{MeV} - 300 \:\text{GeV}& \text{(thick-wall simulation)\,,}
\end{cases}
\end{equation}
which is very similar to the allowed range in $T_*$ shown in Fig.~\ref{fig:spectra}. An ingredient in obtaining these scales
is that the number of e-foldings during the tunneling transition
must satisfy $(H_*/\beta) \sim 1/150- 1/3$ as shown in Fig.~\ref{fig:spectra}.  The upper bound arises from the percolation condition in Eq.~\eqref{eq:percolation}, while the lower bound comes from requiring a large enough normalization of the gravitational wave signal.
In model realizations, the vacuum energy $\rho_{\text{vac}}$ corresponds to the energy density difference between the false and true vacuum in the potential of the tunneling field.

\section{Cosmological Scenarios with a first order Big Bang Phase Transition}\label{sec:scenarios}

Old inflation~\cite{Guth:1980zm} provides the best-known example of a first order phase transition associated with the Big Bang. While the original model fails the bubble percolation condition, simple modifications successfully reheat the universe in a Big Bang phase transition. Furthermore, well-motivated models of the early universe exist, in which the universe becomes vacuum-dominated for a second time after inflation and undergoes a ``late'' Big Bang phase transition. Below we will describe five complementary cosmological scenarios which feature a Big Bang phase transition consistent with the signal observed at NANOGrav, PPTA and EPTA.

\subsection{Double Field Inflation}\label{sec:doublefield}

Double-field inflation~\cite{Adams:1990ds,Linde:1990gz} is a successful model of the early universe, in which inflation ends through a first order phase transition. Just as in old inflation the exponential expansion of space is driven by a scalar field which is initially trapped in a false vacuum. However, double-field inflation evades the empty-universe problem through the inclusion of a second scalar field which introduces a time-dependence in the vacuum decay rate $\Gamma(t)$. At the beginning, $\Gamma$ is suppressed -- thus permitting enough e-folds of inflation -- but later it becomes so large that the phase transition completes rapidly in a Hot Big Bang.
The bubble collisions during the phase transition induce gravitational radiation which can potentially be probed by interferometers and pulsar timing arrays~\cite{Lopez:2013mqa}. We will consider the phase transition which ends inflation as the origin of the NANOGrav, PPTA and EPTA signal (see~\cite{Ashoorioon:2022raz} for a related idea\footnote{In~\cite{Ashoorioon:2022raz} a phase transition after slow-roll inflation is considered as the origin of the NANOGrav signal. Instead, we will focus on the complementary case of double field inflation, in which the phase transition itself terminates the inflationary epoch.}).

From our estimate in Sec.~\ref{sec:properties} it follows that fitting the pulsar timing signals (by a phase transition at the end of inflation) requires an inflation scale $\lesssim 100\:\text{GeV}$. While slow-roll inflation at such a low scale would (typically) require extreme fine-tuning, this is not the case in double-field inflation. The tuning in low-scale slow-roll inflation is linked to the challenge that an extremely flat potential ($M_{\text{P}}\,V'/V\lesssim 10^{-30}$ with $M_{\text{P}}$ denoting the reduced Planck mass) is required for the density fluctuations to match the Cosmic Microwave Background (CMB) amplitude.\footnote{In rolling models, CMB normalization requires 
$A_s = V /(24 \pi^2 M_{\text{P}}^4 \epsilon)
= 2.1 \times 10^{-9}$
so that the slow-roll parameter must satisfy
$\epsilon \equiv (M_{\text{P}}^2/2) (V'/V)^2 \sim 5.7 \times 10^{-76} \, [V / (10\:\text{MeV})^4] . $}
However, in double-field inflation a constant contribution to the potential during inflation is provided by the energy density of the false vacuum. Hence, an effectively very flat potential -- as needed for low-scale inflation -- can more naturally be realized.

\begin{figure}[t]
\begin{center}
\includegraphics[width=0.49\textwidth]{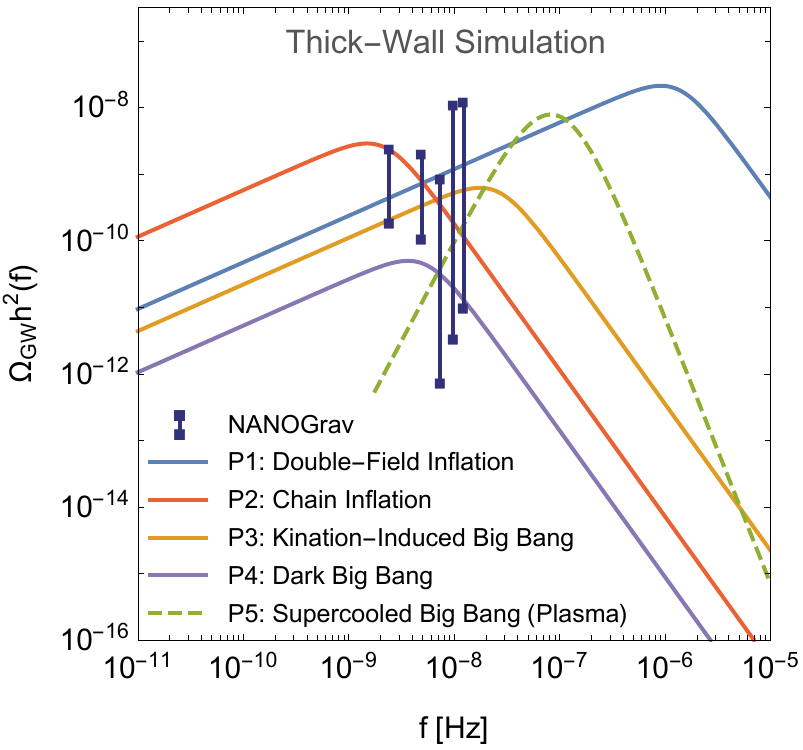}
\includegraphics[width=0.49\textwidth]{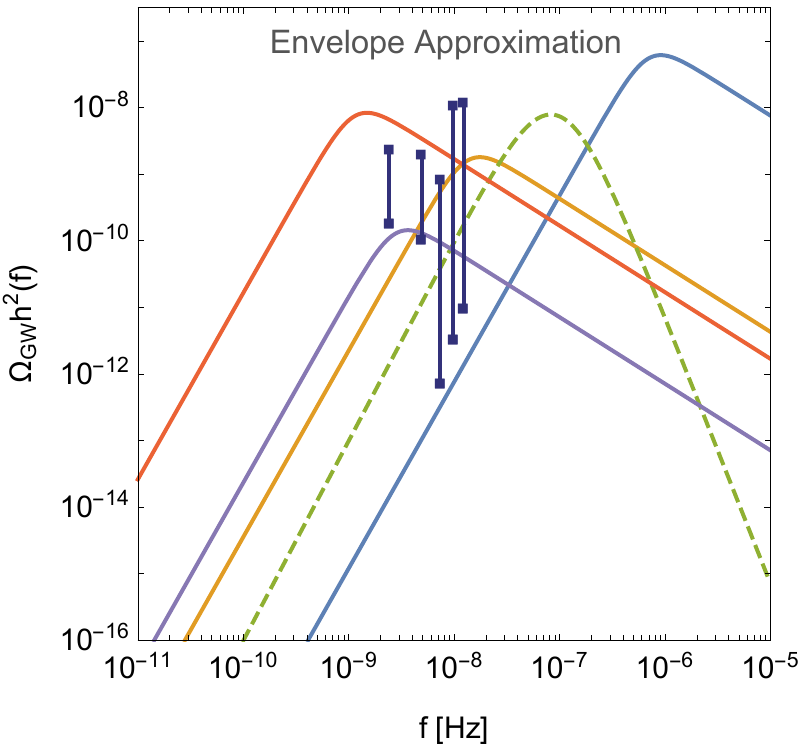}
\end{center}
\caption{Gravitational wave spectra in several cosmological scenarios (P1-P5) with a Big Bang phase transition (as described in Sec.~\ref{sec:scenarios}). The corresponding phase transition parameters can be found in Tab.~\ref{tab:dfiparameters}$\,$-$\,$\ref{tab:dbbparameters}. For the scenarios P1-P4 we depict the spectrum from bubble collisions as predicted by the thick-wall simulation (left panel) and by the envelope approximation (right panel). For the scenario P5 plasma-induced gravitation waves dominate and we, hence, depict the acoustic gravitational wave spectrum (dashed line in both panels).}
\label{fig:spectra2}
\end{figure}

\begin{figure}[htp]
\begin{center}
\includegraphics[width=0.6\textwidth]{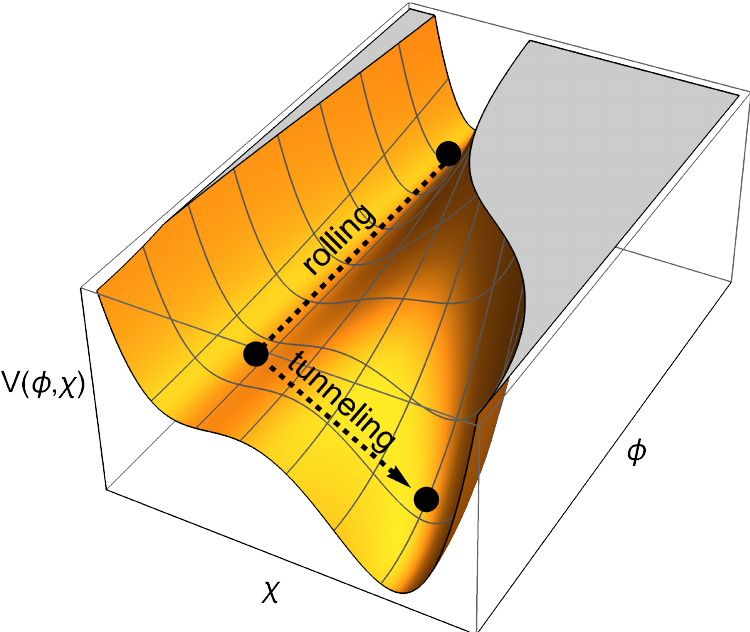}
\end{center}
\vspace{-5mm}
\caption{Potential in double-field inflation. The inflaton $\phi$ is initially displaced far from its minimum. The tunneling field $\chi$ is trapped in a metastable minimum through its coupling to the inflaton. While the inflaton slowly rolls down its potential a second deeper minimum in $\chi$-direction occurs. Once the barrier between the two minima becomes sufficiently small $\chi$ tunnels into the true minimum and inflation ends by a first order phase transition.}
\label{fig:potential3d}
\end{figure}

The basic mechanism of double field inflation is illustrated in Fig.~\ref{fig:potential3d} which depicts the two-field potential. Initially, the inflaton field $\phi$ is displaced far from its minimum. The tunneling field $\chi$ is held in a false vacuum through its coupling to the inflaton. While $\phi$ slowly rolls down its potential, a deeper (=true) minimum in $\chi$-direction appears, and simultaneously the barrier between the two minima becomes shallower. The tunneling rate of $\chi$ into the true minimum increases with time. Once the inflaton reaches a critical field-value $\phi_*$ -- roughly when one true-vacuum bubble is formed per Hubble patch -- inflation ends in a first order phase transition with $\chi$ tunnelling into the true vacuum. The vacuum bubbles collide quickly and reheat the universe successfully.

The idea of double-field inflation can be implemented in a plethora of model realizations. As a simple example, we consider the following two-field Lagrangian,
\begin{equation}\label{eq:doubleL}
\mathcal{L}= \frac{1}{2}\left(1-\frac{\phi^2}{\Lambda^2}\right)^{-2}\partial_\mu \phi\partial^\mu \phi + \frac{1}{2}\partial_\mu \chi\partial^\mu \chi - V(\phi,\chi)\,,
\end{equation}
with
\begin{equation}\label{eq:doubleV}
V(\phi,\chi)= \frac{m_{\phi}^2}{2}\phi^2 + \kappa \phi^2\chi^2 + V_0+\frac{m_\chi^2}{2}\chi^2 - \mu \chi^3 + \lambda^2 \chi^4\,.
\end{equation}
The potential exhibits a metastable minimum with energy density $V_0$ at $\chi=0$, while the global minimum is located at $\phi=0$, $\chi=(3\mu+\sqrt{9\mu^2-16\lambda^2 m_\chi^2})/(8\lambda^2)$.\footnote{We assumed $\mu > 4\lambda m_\varphi/3$.} We chose $V_0$ such that the potential energy vanishes in the true minimum. 

Double field inflation with the potential in Eq.~\eqref{eq:doubleV} but with canonical kinetic terms has previously been discussed in~\cite{Copeland:1994vg,Cortes:2009ej}. Since this minimal realization is now in tension with CMB constraints\footnote{We note that in general models with convex potentials for rolling fields are no longer a good fit to the data.} one needs to slightly modify the original scheme. As a simple possibility we considered in Eq.~\eqref{eq:doubleL} is a non-canonical kinetic term of the inflaton as motivated in the context of $\alpha$-attractor inflation~\cite{Kallosh:2013hoa}.\footnote{The resulting double field inflation model also bears some resemblance to hybrid $\alpha$-attractor inflation recently proposed in~\cite{Kallosh:2022ggf}.}

The following field redefinition allows us to express the Lagrangian in terms of the canonically normalized inflaton field $\hat{\phi}$, 
\begin{equation}
\phi = \Lambda \tanh\left(\frac{\hat{\phi}}{\Lambda}\right)\,.
\end{equation}

During inflation $\chi$ is trapped in the metastable minimum at $\chi=0$ and the potential in inflaton direction (in the canonically normalized basis) is given as
\begin{equation}\label{eq:infpotdf}
V = V_0 +\frac{m_\phi^2 \Lambda^2}{2} \tanh^2\left(\frac{\hat{\phi}}{\Lambda}\right)\,.
\end{equation}
Successful inflation can be realized for any value of $V_0$. However, if $V_0$ is subdominant we would essentially be left with a slow-roll inflation model, which is not the focus of this work. Instead we will concentrate on the regime of ``true'' double-field inflation, where the energy density during inflation is dominated by the false vacuum energy $V_0$.

The inflaton is initially displaced from its minimum, thereby contributing to the effective mass of the tunneling field,
\begin{equation}\label{eq:effmass}
 m_{\chi,\text{eff}}^2 = m_\chi^2 + 2 \kappa \Lambda^2 \tanh^2\left(\frac{\hat{\phi}}{\Lambda}\right)\,.
\end{equation}
For large $\hat{\phi}$ the $\chi$-field is strongly stabilized at $\chi=0$. However, as the inflaton rolls down its potential, a second minimum in $\chi$-direction develops which eventually becomes energetically favorable (see Fig.~\ref{fig:potential3d}). The universe still remains in the false vacuum for some time due to the potential barrier separating the two minima. But eventually $\chi$ tunnels into the true minimum and inflation ends in a first order phase transition.

Given that $V_0$ is dominant compared to all other energy scales in the problem, the vacuum transition occurs mostly in $\chi$-direction. In order to obtain the tunneling rate we can thus employ the analytic approximation for single-field tunneling in a quartic potential~\cite{Adams:1993zs},
\begin{equation}\label{eq:tunneling_doublefield}
\Gamma \simeq m_{\chi,\text{eff}}^4 \left(\frac{S_4}{2\pi}\right)^2 e^{-S_4}\,,\qquad
S_4 = \frac{\pi^2\mu^6}{24\lambda^2(\mu^2-2\lambda^2 m_{\chi,\text{eff}}^2)^3}\sum\limits_{i=1}^3 A_i \left(\frac{\lambda  m_{\chi,\text{eff}}}{\mu}\right)^{2i}\,,
\end{equation}
with $A_1=55.328$, $A_2=- 173.104$ and $A_3=132.896$.

The duration of the phase transition $\beta^{-1}$ is obtained from Eq.~\eqref{eq:beta2}. We can approximate
\begin{equation}\label{eq:betadoublefield}
\beta \simeq \left.\frac{\dot{\Gamma}}{\Gamma}\right|_{t=t_*} 
\simeq -\left.\dot{S_4}\right|_{t=t_*}\simeq \left.\frac{1}{3H}\frac{\partial S_4}{\partial\hat{\phi}}\frac{\partial V}{\partial \hat{\phi}}\right|_{\hat{\phi}=\hat{\phi}_*} \,,
\end{equation}
where we used that the time-dependence of $\Gamma$ dominantly arises from the time-dependence of the Euclidean action of the bounce. Furthermore, we employed the equation of motion $3H\dot{\hat{\phi}}+\partial V/\partial \hat{\phi}\simeq 0$ in the last step.

In order to derive the critical inflaton field-value $\hat{\phi}_*$ at which the tunneling is triggered, we need to determine the time of the phase transition $t_*$. The latter is defined by the condition $I(t_*)=1$ with the integral $I$ from Eq.~\eqref{eq:prob}. We note that $I(t_*)$ is strongly dominated by times around $t_*$. 
Thus we can replace $a(t^\prime) r_{\rm com}(t,t^\prime)$ by $(t-t^\prime)$ in the integral. Furthermore, expanding the bounce action in the exponent of Eq.~\eqref{eq:tunneling_doublefield}
around $t=t_*$ and using Eq.~\eqref{eq:betadoublefield} we approximate $\Gamma(t) \simeq \Gamma(t_*) e^{\beta(t-t_*)}$.  We obtain $I(t_*)=8\pi \Gamma(t_*)/\beta^4 = 1$ and, hence,
\begin{equation}\label{eq:Gammatstar}
\Gamma(t_*) = \frac{\beta^4}{8\pi}\,.
\end{equation}
Plugging Eq.~\eqref{eq:tunneling_doublefield} and~\eqref{eq:betadoublefield} into~Eq.~\eqref{eq:Gammatstar} yields an implicit equation for $\hat{\phi}_*$ which can be solved numerically.

Let us now turn to the cosmological predictions of double field inflation. The perturbations seeding the CMB anisotropies are generated by quantum fluctuations of $\phi$ during the slow-roll phase before the phase transition. Therefore, CMB observables are calculated in the slow-roll formalism. Defining the slow roll parameters,
\begin{equation}
\epsilon=\left.\frac{M_{\text{P}}^2}{2}\left(\frac{\partial V/\partial \hat{\phi}}{V}\right)^2\right|_{\chi=0}\,,
\quad\eta    =\left.M_{\text{P}}^2\,\frac{\partial^2 V/\partial \hat{\phi}^2}{V}\right|_{\chi=0}\,,
\end{equation}
we can employ the standard expressions for the normalization $A_s$ and the spectral index $n_s$ of the scalar power spectrum. Taking into account that the energy density during inflation is dominated by $V_0$, we arrive at,
\begin{equation}\label{eq:observables}
A_s\simeq \left.\frac{V_0}{24\pi^2\,M_{\text{P}}^4\,\epsilon}\right|_{aH=k_{\text{pivot}}}\,,\qquad
n_s\simeq  1 - 6 \epsilon + 2\eta\Big|_{aH=k_{\text{pivot}}}\,,\qquad
r \simeq 16\epsilon\Big|_{aH=k_{\text{pivot}}}
\, .
\end{equation}
The quantities above are evaluated at horizon crossing of the Pivot scale 
$k_{\text{pivot}}=0.05\:\text{Mpc}$ of density fluctuations observable in the CMB.
The number of e-foldings between the horizon crossing of the pivot scale and the end of inflation is given by
\begin{equation}\label{eq:Nk1}
N(k_{\text{pivot}}) = \int\limits_{\hat{\phi}_*}^{\hat{\phi}_{\text{pivot}}} \frac{d\hat{\phi}}{\sqrt{2\epsilon}}\,. 
\end{equation}
where we defined $\hat{\phi}_{\text{pivot}}$ as the field value at which the inflaton resides when $aH=k_{\text{pivot}}$. The critical field value $\hat{\phi}_*$ at which tunneling is triggered determines the end of inflation.
This is in contrast to conventional slow-roll inflation where the lower boundary of the integral (the end of inflation) is set by the field-value at which the slow-roll conditions are violated. $N(k_{\text{pivot}})$ is fixed by the energy scale of inflation
\begin{equation}\label{eq:Nk2}
 N(k_{\text{pivot}}) = \log\left(\frac{a_*}{a_{\text{pivot}}}\right) 
 = \log\left(\frac{a_* H_*}{k_{\text{pivot}}}\right)\simeq 19.2 - \frac{1}{12}\log\left(g_{\text{eff}}(T_*)\right)+ \log \left( \frac{V_0^{1/4}}{\text{GeV}} \right)\,,
\end{equation}
where we denoted the scale factor and Hubble scale at horizon crossing of the Pivot scale during inflation by $a_{\text{pivot}}$ and $H_{\text{pivot}}$. In the second step we approximated $H$ as being constant throughout the epoch of inflation so that $H_{\text{pivot}} \simeq H_*$.  Since $V_0$ dominates the inflaton potential in Eq.(\ref{eq:infpotdf}) (by many of orders of magnitude), 
this approximation is very accurate.
Here $T_*$ is the reheating temperature (= the temperature of the radiation plasma directly after the phase transition),
\begin{equation}
T_* =\left(\frac{30\,V_0}{\pi^2 g_{\text{eff}}(T_*)}\right)^{1/4}\,.
\end{equation}
Since the phase transition completes within a small fraction of a Hubble time, we approximated it as instantaneous for deriving $N(k_{\text{pivot}})$ above. The corresponding error on $N(k_{\text{pivot}})$ is negligible. 
The inflaton-field value $\hat{\phi}_{\text{pivot}}$ can now be obtained by combining Eq.~\eqref{eq:Nk1} and Eq.~\eqref{eq:Nk2}.

\begin{figure}[t]
\begin{center}
\includegraphics[width=0.9\textwidth]{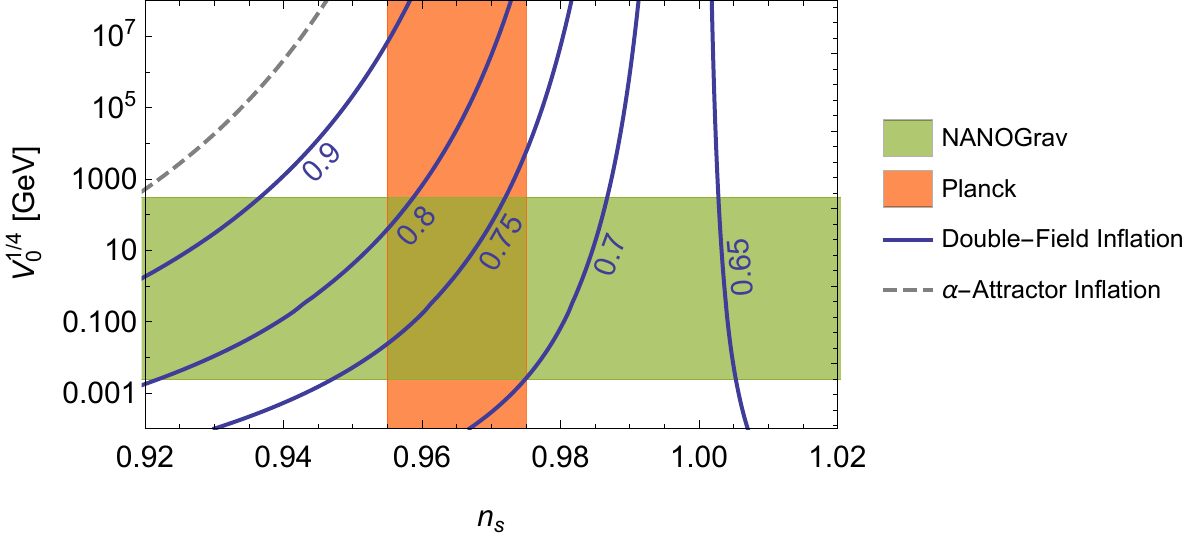}
\end{center}
\vspace{-5mm}
\caption{Spectral index vs.\ false vacuum energy density (=scale of inflation) in the double-field inflation model defined in Eq.~\eqref{eq:doubleL} for different ratios of $\hat{\phi}_{\text{pivot}}/\Lambda$ (as indicated in the figure). Also shown is the range of $V_0$ in which the NANOGrav signal can potentially be explained and the Planck $2\sigma$ constraints on the spectral index~\cite{Planck:2018jri}. For comparison the spectral index in standard $\alpha$-attractor inflation models $n_s=1-2/N(k_{\text{pivot}})$ is also shown by the dashed line (for this case the y-axis corresponds to the scale of inflation; also note that gravitational waves by bubble collisions are not generated in standard $\alpha$-attractor inflation). The $n_s$-prediction of $\alpha$-attractor inflation is approached in the described double field inflation model in the limit $\hat{\phi}_{\text{pivot}}/\Lambda\rightarrow\infty$. The range of inflation scales favored by NANOGrav correspond to a tensor-to-scalar ratio $r=10^{-78}-10^{-56}$ (cf. Eq.~\eqref{eq:r}).}
\label{fig:nsplot}
\end{figure}

The inflaton potential in Eq.~\eqref{eq:infpotdf} is suitable for low-scale inflation required to fit the signal observed by pulsar timing arrays. The plateau in $\hat{\phi}$-direction resulting from the pole in the kinetic term amounts to an inflationary attractor even if the initial energy density of the universe strongly exceeds $V_0$. In order to arrive at a viable model we impose the correct normalization of the scalar power spectrum $A_s=2.1\times 10^{-9}$~\cite{Planck:2018jri} (cf.~Eq.~\eqref{eq:observables}) and the e-fold condition~\eqref{eq:Nk2} which allows us to eliminate $m_\phi$ and $\hat{\phi}_{\text{pivot}}$. The spectral index is then determined by $V_0$ and $\Lambda$ (or more conveniently $\hat{\phi}_{\text{pivot}} / \Lambda$) as depicted in Fig.~\ref{fig:nsplot}.\footnote{The spectral index also exhibits a mild dependence on $\hat{\phi}_*$ which we fixed to $\hat{\phi}_{\text{pivot}}/10$ in Fig.~\ref{fig:nsplot}.} It can be seen that fitting the NANOGrav signal, while simultaneously fulfilling the CMB constraints on $n_s$~\cite{Planck:2018jri}, requires $\hat{\phi}_{\text{pivot}} / \Lambda\simeq 0.7 -0.8$.   

We emphasize, however, that the tensor-to-scalar ratio is highly suppressed in double field inflation models which can fit the pulsar timing signals. By using~Eq.~\eqref{eq:observables} and imposing again the correct normalization of the scalar power spectrum, we obtain
\begin{equation}\label{eq:r}
r \simeq 9.1\times 10^{-67} \:\left(\frac{V_0}{\text{GeV}^4}\right)\,.
\end{equation}
For the range of scales favored by the NANOGrav signal (green band in Fig.~\ref{fig:nsplot}) we find $r=10^{-78}-10^{-56}$. Hence, we can conclude that whenever the gravitational waves from the phase transition at the end of double field inflation are observable (with pulsar timing arrays), tensor modes from inflation are completely negligible.

Fig.~\ref{fig:nsplot} also shows the original rolling $\alpha$-attractor inflation model (dashed line). In this case there is only a single scalar field. For the purposes of this figure we take the y-axis (labeled as $V_0$) to represent the scale of inflation for that model, i.e. only the tanh term in the potential in Eq.(\ref{eq:infpotdf}). One can see that the original slow roll $\alpha$-attractor inflation fails to reproduce the observed $n_s$ in CMB data for potentials at low energy scales.
Further, since it is a slow roll model of inflation, there are no bubbles produced and hence no gravitational waves capable of explaining pulsar timing data.

 On the other hand, $\alpha$-attractor variants at low inflation scales can succeed in two field models.
The double-field model presented here 
 (that uses the non-canonical kinetic term of $\alpha$-attractor models but ends in a first order phase transition) can be successful for potentials of any energy scale, including the range $V_0 \sim$ MeV - 100 GeV that is required by NANOGrav data. Secondly, hybrid $\alpha$-attractor inflation~\cite{Kallosh:2022ggf} can also give rise to low energy inflation, although there are no bubble collisions and hence the model cannot explain pulsar timing data.
In these two field models, the potential in Eq.(\ref{eq:infpotdf}) is dominated by the $V_0$ term set by the second field (the tunneling field in the double field inflation case), a term not
present in single field $\alpha$-attractor inflation.

\begin{table}[htp]
\begin{center}
\begin{tabular}{|ll|ll|}
\hline
&&&\\[-4mm]
\multicolumn{2}{|c|}{Input Parameters}& \multicolumn{2}{c|}{Inflation/ CMB} \\
 \hline
&&&\\[-4mm] 
$\Lambda$~[meV] & $3.2$ & $V^{1/4}(\hat{\phi}_{\text{pivot}})$~[GeV] & $13.7$\\[1mm]
$m_\phi$~[$\mu$eV] & $0.028$ & $N(k_{\text{pivot}})$ & $21.4$\\[1mm]
$m_\chi$~[$\mu$eV] & $36.2$ & $A_s$ & $2.1\times 10^{-9}$\\[1mm]
$\mu$~[$\mu$eV] & $53.6$ & $n_s$ & $0.965$\\[1mm]
$\lambda$ & $1.7\times 10^{-10}$ & $r$ & $3.2\times 10^{-62}$\\
\cline{3-4}
&&&\\[-4mm]
$\kappa$ & $0.02$ & \multicolumn{2}{c|}{Phase Transition} \\
\hline
&&&\\[-4mm]
\multicolumn{2}{|c|}{Derived Parameters} & $T_*$~[GeV] & $5.9$ \\
\cline{1-2}
&&&\\[-4mm]
$\hat{\phi}_{\text{pivot}} / \Lambda$ & $0.76$ & $\beta/H_*$ & $4.7$ \\[1mm]
$\hat{\phi}_* / \hat{\phi}_{\text{pivot}}$ & $0.1$ & $\alpha$ & $\infty$\\
\hline
\end{tabular}
\end{center}
\vspace{-0.4cm}
\caption{Benchmark point for the double field model containing the inflaton $\phi$ and the tunneling field $\chi$ as defined in Eq.~\eqref{eq:doubleL}. The benchmark point can explain the tentative gravitational wave signal observed at several pulsar timing arrays. Input parameters and predictions for the scale and e-foldings of inflation, CMB observables and phase transition parameters are shown. The corresponding gravitational wave spectrum is depicted in Fig.~\ref{fig:spectra2} (labelled by P1).}
\label{tab:dfiparameters}
\end{table}

In Tab.~\ref{tab:dfiparameters} we provide a benchmark point for successful double field inflation ending in a Big Bang phase transition. The corresponding gravitational wave spectrum is shown together with the NANOGrav data in Fig.~\ref{fig:spectra2} (labelled by P1). The location of the benchmark point in the thick-wall regime of vacuum tunnelling suggests to employ the spectrum of the thick-wall simulation (left panel of the figure). As can be seen a good fit to the NANOGrav signal is obtained. Since only the lower tail of the signal falls into the frequency band of pulsar timing arrays, the measured spectrum is well described by a single power law with amplitude $\overline{\Omega}=3\times 10^{-9}$ and index $\bar{\gamma}=0.7$. Fig~\ref{fig:nanogravPL} immediately reveals that such a power law spectrum also well describes the PPTA and EPTA data. Hence, double field inflation is a good candidate for generating the tentative gravitational wave signal observed by the pulsar timing arrays.

\subsection{Chain Inflation}\label{sec:chaininflation}

Chain Inflation~\cite{Freese:2004vs,Freese:2005kt,Ashoorioon:2008pj} is another well-motivated model of the early universe with a first order phase-transition origin of matter and radiation. In contrast to old inflation, chain inflation features a series of consecutive first order phase transitions instead of a single one. Each individual transition proceeds rapidly within a small fraction of a Hubble time such that the bubble percolation condition is easily satisfied. And yet -- due to the presence of many individual vacua -- inflation can easily last for the $15-60$ e-folds required to resolve the horizon problem. 

Radiation, matter and gravity waves are generated at each of the phase transitions along the chain -- there are thus many consecutive Hot Big Bangs. However, since matter and radiation produced early during inflation are quickly redshifted away, it is the last few Big Bangs which generate the energy content observed in our present universe.

In order to fit the pulsar timing array signals in chain inflation we will again be drawn to a low inflation scale in the sub-TeV regime. In this light, it is important to point out that low-scale chain inflation can be realized without parameter tuning. This is different from low-scale slow roll inflation which requires an extremely flat (typically tuned) potential in order to match the observed CMB amplitude. The advantage of chain inflation arises due to the origin of the CMB anisotropies which (in contrast to slow roll inflation) is not linked to quantum fluctuations of the inflaton -- the latter are suppressed by the inflaton mass in each of the vacua. Rather, the probabilistic nature of tunneling -- different patches of the universe undergo tunneling at slightly different times -- causes density perturbations in the primordial plasma which later manifest as the anisotropies in the CMB. As we will see below, the CMB amplitude in chain inflation is determined by the tunneling rate normalized to the Hubble rate. Hence, no particular requirements on the flatness of the potential arise in low-scale chain inflation.

The CMB observables of chain inflation have recently been derived through dedicated simulations in~\cite{Winkler:2020ape}. In the following, we denote the vacuum in which the universe resides during horizon crossing of the the Pivot scale of the CMB   by $n=0$, the next vacuum in the chain by $n=1$, the next-to-next vacuum by $n=2$ and so on. An index $n$ indicates that a quantity is evaluated in the $n$th vacuum. 
This choice of definition of $n=0$ at horizon crossing of the pivot scale has been made for convenience of notation, since this is the scale at which $n_s$ and $r$ are determined from CMB observations. We note, however, that chain inflation began earlier, with the vacuum residing in higher values of the potential; i.e. in the current notation chain inflation began already at negative values of $n$.  Indeed CMB observables at the largest length scales arise from these earlier phase transitions, which would be relevant for determining e.g the running of the spectral index.

In our notation, the scalar power spectrum and the scalar spectral index are given as
\begin{equation}\label{eq:chainobservables}
A_s \simeq 0.06 \left(\frac{\Gamma_0^{1/4}}{H_0} \right)^{-5/3}\,,\qquad
n_s \simeq 1+ 0.58\,\left(\frac{\Gamma_0^{1/4}}{H_0} \right)\,\left( \frac{2\Delta V_0}{V_0} - \frac{\Delta \Gamma_0}{\Gamma_0} \right)\,,
\end{equation}
where $\Delta \Gamma_n=\Gamma_{n+1}-\Gamma_n$ and $\Delta V_n=V_{n+1}-V_n$. Note that in the above expression $H_0$ stands for the Hubble rate in the $0$th vacuum and not for the Hubble constant today.

\begin{figure}[htp]
\begin{center}
\includegraphics[width=0.5\textwidth]{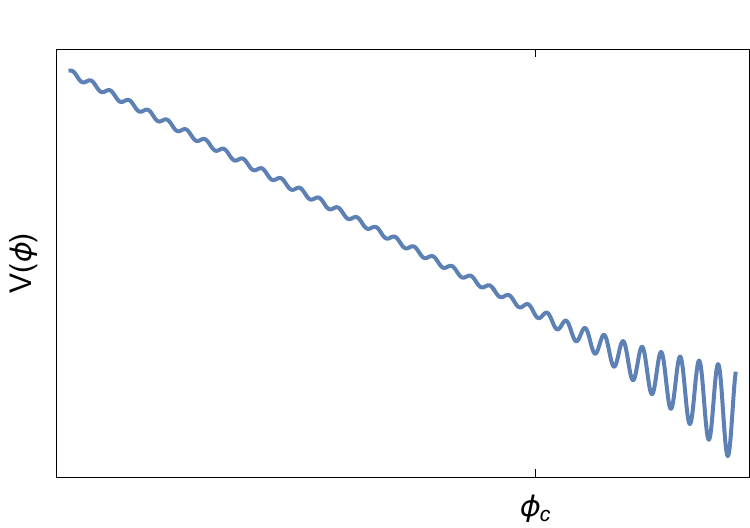}
\end{center}
\vspace{-5mm}
\caption{Illustration of the inflaton potential in the chain inflation model defined by Eq.~\eqref{eq:chainmodel}. During inflation $\phi$ tunnels from minimum to minimum along the tilted cosine potential. Once it reaches the critical field value $\phi_c$ the stopping mechanism is triggered and the potential barriers increase quickly, thereby stopping the inflaton in a quasistable minimum (see text).}
\label{fig:chainplot}
\end{figure}

A prime candidate for the inflaton in chain inflation is an axion in a quasi-periodic potential. We consider the following simple realization
\begin{equation}\label{eq:chainmodel}
 V = \Lambda^4 \cos\left( \frac{\phi}{f} \right) - \mu^3 \phi  + V_{\text{stop}}\,,
\end{equation}
where the parameters $\Lambda$, $\mu$ and $f$ are chosen such that the potential exhibits a series of metastable minima (which implies $\Lambda^4 > f\mu^3$). During inflation $\phi$ tunnels along the minima of the tilted cosine. The last term is irrelevant for tunneling during inflation, but ensures that the inflaton stops in a (quasi)stable minimum once the vacuum energy has been dissipated. A possible choice -- familiar from the relaxion mechanism~\cite{Graham:2015cka} -- is\footnote{This model has recently been considered as a realization of early dark energy~\cite{Freese:2021rjq}.}
\begin{equation}\label{eq:relax}
 V_{\text{stop}} = (M_1^2 - M_2 \phi)\chi^2 + {\Lambda^\prime}^2 \chi^2\cos\frac{\phi}{f} + \lambda\chi^4+\text{const} \,.
\end{equation}
The auxiliary field $\chi$ is initially stabilized at $\chi=0$ and decouples from inflation. But once the inflaton passes the critical field value $\phi_c \simeq M_1^2/M_2$ the $\chi$-field gets displaced. Thereby it raises the potential barriers in $\phi$-direction and quickly stops the tunneling in a minimum with vanishing vacuum energy (the latter is ensured through appropriate choice of the constant in Eq.~\eqref{eq:relax}). The inflaton potential with $\chi$ set to its minimum is depicted in Fig.~\ref{fig:chainplot}.

At $\phi<\phi_c$ the inflaton potential is a pure tilted cosine and the tunneling rate remains constant.\footnote{This strictly holds if temperature effects on the tunneling rate can be neglected which is justified if the coupling between $\phi$ and the radiation generated by earlier phase transitions is sufficiently suppressed. We emphasize, however, that the absence of temperature effects on $\Gamma$ is not crucial for realizing chain inflation. We merely avoided the additional model-dependence in the presence of temperature effects for the sake of a simple discussion.} But once the stopping mechanism is triggered, the tunneling rate decreases exponentially due to the exponential dependence of $\Gamma$ on the Euclidean action of the bounce. (cf.\ Eq.~\eqref{eq:tunnelingrate}). We can parametrize $\Gamma_n$ in the following way\footnote{The Euclidean action of the bounce scales approximately as $S_4\propto (\Lambda^4+{\Lambda^\prime}^2 \chi^2 )^2 $ for the stopping potential in Eq.~\eqref{eq:relax}. Expanding $S_4$ around $\chi=0$ and taking into account $\chi\propto (n-n_c)$ suggests a quadratic dependence $S_4= S_{4,0} + S^\prime (n-n_c)^2$ on $n$ after the inflaton passes the critical field value.},
\begin{equation}\label{eq:Gamma0}
\Gamma_n = \begin{cases}
\Gamma_0 &\;\; n\leq n_c\,,\\
\Gamma_0 \, e^{-S^\prime (n-n_c)^2} &\;\; n> n_c\,,
\end{cases}
\end{equation}
where $n_c$ is the number of the vacuum in which the stopping mechanism is triggered, i.e.\ the vacuum corresponding to the field value $\phi_c$. If $\phi_c$ lies between two minima in the potential $n_c$ becomes a non-integer number (e.g.\ $n_c=1000.5$ if $\phi_c$ lies in the middle between the 1000th and 1001th minimum of the potential)\footnote{We note that the field always resides in a minimum of the potential, corresponding to an integer value of $n$. However, the tunneling rate may be different at two adjacent minima due to the fact that $\phi_c$ lies in between these two minima.}.

The parameters specifying the tunneling rate in Eq.~\eqref{eq:Gamma0} can easily be linked to the potential parameters through the expressions for the tunneling rate in a quasi-periodic potential as given in Ref.~\cite{Winkler:2020ape}. In the following -- due to the absence of strong theoretical priors on the potential parameters -- we avoid this step and simply define our chain inflation model in terms of $\Gamma_0$, $S^\prime$, $n_c$ and the scale of inflation $V_0$. This choice is most convenient for the comparison with observation. 

After imposing the correct normalization of the power spectrum $A_s=2.1\times 10^{-9}$ we use Eq.~\eqref{eq:chainobservables} to relate the spectral index to the total number of transitions $n_{\text{tot}}$ after horizon crossing of the pivot scale in the CMB $n_{\text{tot}} \simeq V_0/\Delta V_0$. Since only a very small number of transitions occurs after the inflaton passes the critical field-value we can set $n_{\text{tot}}\simeq n_c$. We, hence, obtain
\begin{equation}\label{eq:nc}
n_c = \frac{3.45\times 10^4}{1-n_s} \simeq (0.8-1.3)\times 10^6\,.
\end{equation}
In the last step we imposed the Planck $2\sigma$-constraint $n_s=0.956-0.973$~\cite{Planck:2018jri}. CMB constraints thus require chain inflation to feature a relatively large number of vacuum transitions in the range of $10^6$ (as was previously noted in~\cite{Winkler:2020ape,Freese:2021noj}). 

\begin{figure}[t]
\begin{center}
\includegraphics[width=0.6\textwidth]{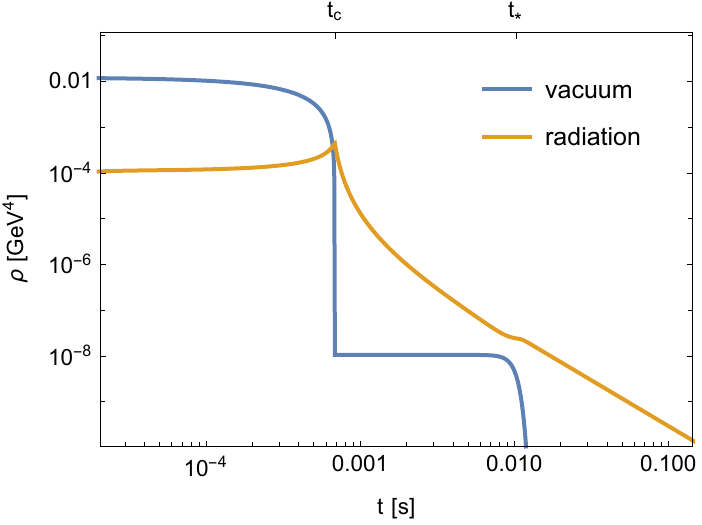}
\end{center}
\vspace{-5mm}
\caption{
Time-evolution of the vacuum and radiation energy densities in the chain inflation model defined in Eq.~\eqref{eq:Gamma0} for the parameter choice of Tab.~\ref{tab:chainiparameters}. During the inflationary epoch ($t\lesssim 5\times 10^{-4}\:\text{s}$) the universe is vacuum-dominated. However, the successive phase transitions reduce the vacuum energy and build up a radiation plasma which eventually starts to dominate the energy budget at $t\sim 5\times 10^{-4}\:\text{s}$. The stopping mechanism is triggered at time $t_c$, at which point the barrier height in the potential increase quickly so that the tunneling rate becomes much slower.  Since the last phase transition (before the inflaton settles) is delayed (see text) a second era of vacuum-significance occurs around $t\sim 0.01\:\text{s}$. The final Hot Big Bang phase transition at time $t_c$ converts the remaining vacuum energy into radiation and releases a strong gravitational wave signal.}
\label{fig:rhochain}
\end{figure}

The number of e-foldings between horizon crossing of the pivot scale in the CMB and the time $t_c$ when the stopping mechanism is triggered (at $\phi=\phi_c$) is given as~\cite{Winkler:2020ape}
\begin{equation}\label{eq:Nkchain}
 N_c \simeq 0.7 \sum\limits_{n=1}^{n_c} \frac{H_n}{\Gamma_n^{1/4}}\simeq  2.4\times 10^{-5} \sum\limits_{n=1}^{n_c}\sqrt{\frac{V_n}{V_0}}\simeq 2.4\times 10^{-5} \sum\limits_{n=1}^{n_c} \sqrt{1-\frac{n}{n_c}}
\simeq 16\,\left(\frac{n_c}{10^6}\right)\,,
\end{equation}
where we again employed the normalization of the scalar power spectrum. Furthermore, we neglected the radiation contribution to $H_n$. A refined estimate taking into account the radiation plasma increases $N_c$ by $\sim 1/2$ compared to Eq.~\eqref{eq:Nkchain}. Plugging Eq.~\eqref{eq:nc} into Eq.~\eqref{eq:Nkchain} and including the small correction yields
\begin{equation}\label{eq:Nc}
N_c =13-21\,,
\end{equation}
for the spectral index in the Planck-observed range. The number $N_c$ is very similar but not identical to the total number of e-folds during observable inflation $N(k_{\text{pivot}})$. As a reminder, we take $N(k_{\text{pivot}})$ to be the number of e-folds between the horizon crossing of the CMB pivot scale and the end of inflation i.e. the onset of the radiation-dominated epoch (shown in Fig.~\ref{fig:rhochain} at
the point where the vacuum (blue) and radiation (yellow) lines cross). On the other hand we take $N_c$ to be the number of e-folds between the horizon crossing of the CMB pivot scale and the trigger of the stopping mechanism at $t_c$ (shown in Fig.~\ref{fig:rhochain} at the point where the vacuum (blue) line takes a 90$^{\circ}$ turn).  During chain inflation a radiation background with energy density $\rho_r\sim V_0/N_c$ is present since it takes $\sim 1$ e-fold to redshift away radiation from earlier phase transitions. Hence, radiation domination starts about one Hubble time before $t_c$. 
The temperature of the universe $T_c$ at the time $t_c$ can be obtained by summing up the contributions to the radiation density  from all previous phase transitions (taking into account their redshift). We find
\begin{equation}\label{eq:Tcchain}
T_c = \left(\frac{30}{\pi^2} \frac{\rho_r(T_c)}{g_{\text{eff}}(T_c)}\right)^{1/4}\quad\text{with}\quad
\rho_r(T_c)\simeq 0.7\, \frac{V_0}{N_c}\,.
\end{equation}
After $t_c$ the universe undergoes a small number of ever slower vacuum transitions before it settles in a quasistable vacuum for its remaining lifetime. While the universe is (strongly) radiation-dominated at $t_c$ it may become vacuum-dominated for a second time if the last vacuum transition occurs sufficiently late. The second vacuum-domination -- if it occurs -- can only last a fraction of an e-fold since the percolation condition would otherwise be violated (reintroducing the empty universe problem of old inflation). For relating the number of e-folds to the scale of inflation we can, hence, neglect this small episode and obtain an expression similar to Eq.~\eqref{eq:Nk2},
\begin{equation}\label{eq:NcV0}
N_c =\log\left(\frac{a_c H_c}{k_{\text{pivot}}}\right)\simeq 19.2 + \log \left( \frac{\rho_r^{1/4}(T_c)}{g_{\text{eff}}^{1/12}(T_c)\:\text{GeV}} \right)\simeq 19.1 + \log \left( \frac{V_0^{1/4}}{N_c^{1/4}\,g_{\text{eff}}^{1/12}(T_c)\:\text{GeV}} \right)\,,
\end{equation}
where we used Eq.~\eqref{eq:Tcchain} in the last step. As previously found, the correct normalization and spectral index of the scalar power spectrum imposes $N_c=13-21$. The corresponding scale of inflation derived from Eq.~\eqref{eq:NcV0} is
\begin{equation}\label{eq:V0range}
 V_0^{1/4} = 5\:\text{MeV}- 20\:\text{GeV}\,.
\end{equation}
We can thus conclude that the simple chain inflation model defined in Eq.~\eqref{eq:chainmodel} is consistent with all cosmological constraints for a low inflation scale in the MeV-GeV-range. Such a low inflation scale immediately suggests a gravitational wave signal in the frequency band of pulsar timing arrays. 

In Fig.~\ref{fig:rhochain} we depict the evolution of the vacuum and radiation densities for the benchmark parameter point in Tab.~\ref{tab:chainiparameters}. It can be seen that the vacuum energy initially dominates and drives the rapid expansion of space. Before the inflaton reaches the critical field value at $t=t_c$, the density $\rho_{\text{vac}}$ decreases linearly with time due to the constant tunneling rate (which looks almost like a step-function in the figure due to the log-log-scale). After $t_c$ the barriers in the inflaton potential start to increase, thereby stopping the tunneling after a few more transitions in a vacuum whose lifetime exceeds the age of the universe. For the benchmark point only two transitions occur after $t_c$. The first one is too close to $t_c$ to be resolved in the figure, whereas the last transition causes the second step in $\rho_{\text{vac}}$ at the time $t_*\simeq 0.01\:\text{s}$. 

\begin{table}[htp]
\begin{center}
\begin{tabular}{|ll|ll|ll|}
\hline
&&&&&\\[-4mm]
\multicolumn{2}{|c|}{Input Parameters}& \multicolumn{2}{c|}{Inflation/ CMB}& \multicolumn{2}{c|}{Phase Transition} \\
 \hline
&&&&&\\[-4mm] 
$\Gamma_0^{1/4}$ & $1.2\times 10^9\:\text{s}^{-1}$ & $V^{1/4}(\phi_{\text{pivot}})$~[GeV] & $0.33$& $T_*$~[MeV] & $9.4$\\[1mm]
$n_c$ & $1.1\times 10^6$ & $N(k_{\text{pivot}})$ & $18.5$& $\beta/H_*$ & $6.7$\\[1mm]
$S^\prime$ & $63.5$ & $A_s$ & $2.1\times 10^{-9}$& $\alpha$ & $0.6$\\[1mm]
 &  & $n_s$ & $0.969$& &\\
\hline
\end{tabular}
\end{center}
\vspace{-0.4cm}
\caption{Chain inflation model parameters entering Eq.~\eqref{eq:Gamma0} and predictions for inflation, CMB observables. Also given are the parameters characterizing the final Hot Big Bang phase transition. The corresponding gravitational wave spectrum (P2 in Fig.~\ref{fig:spectra2}) is consistent with the NANOGrav signal.}
\label{tab:chainiparameters}
\end{table}

Each phase transition along the chain generates new vacuum bubbles which seed radiation and gravity waves upon collision. Also shown in Fig.~\ref{fig:rhochain} is the radiation density which remains approximately constant at $\rho_{\text{r}}\sim V_0/N_c$ during inflation. This is because the continuous increase of $\rho_{\text{r}}$ by bubble collisions cancels with the decrease of $\rho_{\text{r}}$ by redshifting. Shortly before $t_c$ the vacuum energy drops below $V_0/N_c$ and the universe becomes radiation-dominated. However, since the transitions after $t_c$ -- in particular the last one -- take longer, a second era of vacuum significance occurs (falling slightly short of vacuum-domination). This era ends by the last vacuum transition, where $\phi$ tunnels into its present minimum. Until today, the universe stays in this vacuum and evolves according to the cosmological standard model. The time of the final transition $t_*$ is set by $S^\prime$ which determines how quickly the life-time increases from vacuum to vacuum after the stopping mechanism is triggered (cf.\ Eq.~\eqref{eq:Gamma0}).

As noted, in chain inflation there is a large number of Hot Big Bangs in the sense that all phase transitions can contribute to the radiation and matter density of the universe. However it is the last Big Bang phase transition which yields the largest contribution to today's radiation density since it is the least affected by redshifting. The gravitational wave signal for the evolution shown in Fig.~\ref{fig:rhochain} is even entirely dominated by the last phase transition. 
This can be understood due to two factors in Eq.(~\eqref{eq:gravityspectrum}) for  the gravitational wave amplitude.
First, $\Omega_{GW} \propto \bigl( {H_* / \beta} \bigr)^2$, i.e. the gravitational wave amplitude decreases as the square of the inverse of the number of phase transitions per e-fold.
During most of the phase transitions in chain inflation with a tilted cosine, we have seen that matching CMB data requires 
$10^6$ transitions per e-fold, leading to strong suppression ($10^{-12}$) of the gravitational wave amplitude.  Only in the last phase transition, which is much slower due to the stopping mechanism, is there a substantial gravitational wave amplitude produced.
Secondly, the gravitational wave amplitude scales with the fraction of the total energy participating in the phase transition which is maximized for the last transition.

Concentrating on the gravity waves from the last Big Bang, the problem effectively reduced to a single-phase-transition case with a radiation background as discussed in Sec.~\ref{sec:implications}. In order to derive the gravitational wave spectrum we simply need to determine the vacuum and radiation densities $\rho_{\text{vac}}$, $\rho_r(T_n)$ right before the phase transition. While $\rho_{\text{vac}}\simeq V_0/n_c$, $\rho_r(T_n)$ is obtained by adding the contributions from previous phase transitions taking into account their redshift (see Fig.~\ref{fig:rhochain}). The ratio $\alpha = \rho_{\text{vac}}/\rho_r(T_n)$ then also fixes the duration of the phase transition $H_*/\beta$ (see Fig.~\ref{fig:betaalpha}) and the reheating temperature $T_*$ through Eq.~\eqref{eq:Tst}. The corresponding gravitational wave spectrum follows from Eq.~\eqref{eq:gravityspectrum}.

The gravitational wave spectrum for the benchmark point in Tab.~\ref{tab:chainiparameters} is shown in Fig.~\ref{fig:spectra2} for the thick-wall simulation and the envelope approximation (P2 in the figure). For both cases, the predicted spectrum is compatible with the NANOGrav signal.\footnote{In the envelope approximation the amplitude of the predicted spectrum is slightly above the NANOGrav measurement.} As can be seen in Fig.~\ref{fig:spectra} the temperature of the benchmark point is slightly below the PPTA-preferred region. We have checked, however, that a good fit to PPTA can be obtained if one e.g.\ increases the scale of inflation by a factor of a few compared to the benchmark point. Hence, we can conclude that the last Big Bang phase transition at the end of chain inflation could well be the origin of the tentative stochastic gravitational wave background seen at pulsar timing arrays.

More generally, other variants of chain inflation could also produce gravitational waves consistent with the pulsar timing array data. As described above, in this paper we have considered the case of a constant (time-independent) tunneling rate (here via a tilted cosine potential), together with a relaxion stopping mechanism that slows the tunneling down. Another alternative, that could also explain the pulsar timing signals, would be the case of a potential that gives rise to a time-dependent tunneling rate $\Gamma \equiv \Gamma(t)$, but this latter case is the purview of future work.

\subsection{Kination-Induced Big Bang}\label{sec:kinBB}

In a standard cosmological evolution, the universe enters the radiation-dominated era once inflation has completed. However, there also exist well-motivated alternative cosmologies, in which the expansion history is altered. A prime example of non-standard evolution is an epoch of kination in which the universe is dominated by the kinetic energy of a scalar field.
Here we propose a model of ``Kination-Induced Big Bang'', in which a period of kination domination ends via a first order
phase transition that reheats into the ordinary radiation-dominated early history of our universe.

\begin{figure}[htp]
\begin{center}
\includegraphics[width=0.5\textwidth]{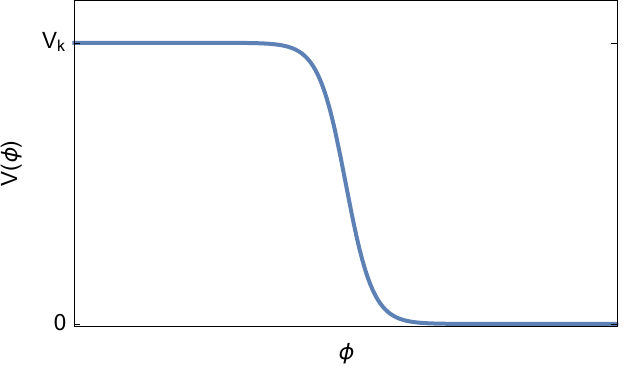}
\end{center}
\vspace{-5mm}
\caption{Potential in quintessential inflation. The inflationary (quintessential) energy density is shown on the left side (right side) of the figure.}
\label{fig:quintV}
\end{figure}

The occurrence of kination is predicted, for example, in models of quintessential inflation which offer a unified explanation of inflation and dark energy. Since the inflationary expansion is driven by the potential energy of a scalar field, there has long been speculation that the same could be true for the accelerated expansion of our present universe. Scalar field models of dark energy -- which predict small deviation of the dark energy equation-of-state parameter from $w=-1$ -- go under the name of quintessence~\cite{Wetterich:1987fm,Ratra:1987rm,Caldwell:1997ii}. The idea of quintessential inflation is to unify the description of inflation and quintessence dark energy in terms of a single scalar field $\phi$. The required potential $V(\phi)$ is depicted in Fig.~\ref{fig:quintV}. Inflation occurs while $\phi$ slowly rolls along the plateau on the left side of the figure. Once it reaches the steeper part of the potential inflation ends and the potential energy is converted into kinetic energy of $\phi$. In contrast to standard slow-roll inflation, $\phi$ does not oscillate around a minimum, but rather continues to `shoot' along the flat bottom of the potential on the right side of the figure. The universe becomes kinetic-energy dominated for some time. However, the kinetic energy redshifts quickly with the sixth power of the scale factor and therefore eventually becomes subdominant to other forms of energy existing in the universe~\cite{Spokoiny:1993kt}. This is when the kination epoch ends. Much later, when (virtually) all kinetic energy has been dissipated, the (tiny) potential energy of $\phi$ once again dominates the energy content of the universe commencing the era of quintessence.

In the following, we will assume that the universe went through an epoch of kination. While we consider quintessential inflation as a prime motivation for kination, let us note that the following discussion can apply to any cosmological scenario running through a kination phase.

A common assumption in kination cosmologies is that the Hot Big Bang occurs prior to the kination phase. If reheating is caused by gravitational particle production at the end of inflation~\cite{Ford:1986sy}, the so-produced plasma is initially subdominant to the kinetic energy, but dominates at a later time due to its slower redshift. However, gravitational reheating has been found to be too inefficient to comply with BBN constraints (see e.g.~\cite{Figueroa:2018twl}) and alternative more complicated mechanisms have been considered.

In this section we propose a new model which we call a Kination-Induced Big Bang. Here, a first order phase transition triggered by the kination field $\phi$ is able to successfully reheat the universe after the kination stage. This new model provides a mechanism for successful reheating that was hard to achieve via gravitational particle production in quintessential inflation,
but (as mentioned above) applies more generally to any cosmological scenario with a kination period. A kination-induced Hot Big Bang can be realized through a derivative coupling of $\phi$ to an auxiliary scalar $\chi$ (=the tunneling field). Such a derivative coupling is particularly attractive if the kination field is identified with the quintessence field in the late universe (as in quintessential inflation), since it avoids strong fifth-force constraints on quintessence. 

We consider the effective two-field Lagrangian,
\begin{equation}\label{eq:Lquint}
\mathcal{L} = \left( \frac{1}{2} + \frac{\chi^2}{M^2} \right)\partial_\mu \phi\partial^\mu \phi
+\frac{1}{2} \partial_\mu \chi\partial^\mu \chi - V(\phi) -\frac{m_\chi^2}{2}\chi^2 + \mu \chi^3 - \lambda^2 \chi^4 + V_0\,,
\end{equation}
valid below the scale $M$. Since we are mostly interested in the Hot Big Bang phase transition after kination, it is sufficient to describe this period in the effective theory given here (there is no need to consider an explicit ultraviolet completion). As initial condition we can set $\dot{\phi}=M^2$ and then follow the evolution of $\dot{\phi}$ through its equation of motion. During kination we can neglect $V(\phi)$ since the potential energy of the kination field must be subdominant (otherwise it would not be kination).

The potential of the auxiliary field features a metastable minimum at $\chi=0$ and a global minimum at $\chi=(3\mu+\sqrt{9\mu^2-16\lambda^2 m_\chi^2})/(8\lambda^2)$. We fix $V_0$ such that the potential energy of $\chi$ vanishes in the true minimum. For this choice, $V_0$ is equal to the false vacuum energy density. During kination the coupling between $\chi$ and $\partial_\mu \phi$ increases the effective mass of the auxiliary field,
\begin{equation}\label{eq:mchikination}
 m_{\chi,\text{eff}}^2 = m_\chi^2 + 2 \frac{\dot{\phi}^2}{M^2}\,,
\end{equation}
which stabilizes $\chi$ in the metastable minimum. This bears resemblance to double field inflation, where a direct coupling to the inflaton was used to stabilize the auxiliary field in a false vacuum (see Sec.~\ref{sec:doublefield}). However, we emphasize that the mechanism described above does not operate during inflation, but rather during the kination stage. 

At the beginning of kination -- when $\dot{\phi}$ is maximal -- the minimum at $\chi=0$ is energetically favorable due to the large effective mass of the auxiliary field. Even if $\chi$ was displaced during inflation it quickly settles in this minimum once kination starts. Subsequently, the Hubble friction reduces $\dot{\phi}$, $m_{\chi,\text{eff}}$ and the second deeper minimum at $\chi\neq 0$ starts showing up. For some time, the universe still remains in a metastable state until $\dot{\phi}$ falls below a critical value $\dot{\phi}_c$ at which the universe tunnels into the true minimum of $\chi$. This critical moment $t_*$ is defined in terms of the tunneling rate by Eq.~\eqref{eq:Gammatstar}, where the tunneling rate is determined by Eq.~\eqref{eq:tunneling_doublefield} (with $m_{\chi,\text{eff}}$ taken from Eq.~\eqref{eq:mchikination}). The phase transition leads to the formation of true vacuum bubbles which collide and reheat the universe in a Hot Big Bang. The energy density of the universe at the time of the phase-transition can be estimated as
\begin{equation}
\rho_{\text{tot}}(t_*) = \frac{\dot{\phi}_c^2}{2} + V_0\,.
\end{equation}
We note that -- depending on the parameter choice -- the phase transition may occur during kination or shortly after kination. In the second case, the universe undergoes a second period of vacuum domination driven by $V_0$. The second vacuum-domination -- if it occurs -- can, however, only last very briefly. Otherwise $\dot{\phi}$ would be completely redshifted away making it implausible that the evolution of $\dot{\phi}$ triggers the phase transition.

Finally, the duration of the phase transition $\beta^{-1}$ is given as (cf. Eq.~\eqref{eq:beta2}),
\begin{equation}
\beta \simeq \left.\frac{\dot{\Gamma}}{\Gamma}\right|_{t=t_*} 
\simeq -\left.\dot{S_4}\right|_{t=t_*}\simeq \left. 3H\dot{\phi}\,\frac{\partial S_4}{\partial\dot{\phi}}\right|_{\dot{\phi}=\dot{\phi}_c}\,,
\end{equation}
with $S_4$ again taken from Eq.~\eqref{eq:tunneling_doublefield}. In the last step we employed the equation of motion $\ddot{\phi} + 3 H \dot{\phi}\simeq 0$, where we used that $V(\phi)$ is negligible at the time of the phase transition. Only in the late universe, long after the Big Bang phase transition, $V(\phi)$ starts to become important again (potentially playing the role of dark energy if $\phi$ is identified with the quintessence field).

\begin{table}[htp]
\begin{center}
\begin{tabular}{|ll|ll|}
\hline
&&&\\[-4mm]
\multicolumn{2}{|c|}{Input Parameters}& \multicolumn{2}{c|}{Phase Transition} \\
 \hline
&&&\\[-4mm] 
$M$~[GeV] & $1.8$ & $V_0^{1/4}$~[MeV] & $27.3$\\[1mm]
$m_\chi$~[MeV] & $0.12$ & $T_*$~[MeV] & $19.6$\\[1mm]
$\mu$~[MeV] & $0.048$ & $\beta/H_*$ & $38.4$\\[1mm]
$\lambda$ & $0.01$ & $\alpha$ & $1070$\\
\hline
\end{tabular}
\end{center}
\vspace{-0.4cm}
\caption{Parameter example yielding a kination-induced Big Bang consistent with the gravitational wave signal at several pulsar timing arrays. Input parameters entering the Lagrangian defined in Eq.~\eqref{eq:Lquint} are shown on the left side, predictions for the phase transition parameters on the right side ($\alpha$ in this case was defined as the ratio of vacuum energy to kinetic energy). The gravitational wave spectrum is shown in Fig.~\ref{fig:spectra2} (labelled by P3).}
\label{tab:quparameters}
\end{table}

The gravitational wave spectrum from the phase transition is derived from Eq.~\eqref{eq:gravityspectrum}. In Tab~\ref{tab:quparameters} we provide a parameter example for a kination-induced Big Bang. The corresponding gravitational wave signal is depicted in Fig.~\ref{fig:spectra2} (labelled by P3). As can be seen, a good fit to the signal of the NANOGrav experiment is obtained. The same is true for PPTA as visible in Fig.~\ref{fig:spectra}, where the benchmark point is also indicated by P3.

We also wish to point out that our model, of kination ending in a first order phase transition, can be used as a mechanism to reheat Quintessential Inflation.

\subsection{Supercooled Big Bang}\label{sec:supercooledBB}

The Hot Big Bang is commonly identified with the reheating process at the end of inflation. However, there exist attractive cosmological scenarios in which the early universe went through a second (short) period of vacuum domination, lasting for e.g. 1 - 10 e-folds. Whereas the earlier epoch of inflation is required to solve the cosmological and horizon problems as
well as generate the density perturbations for the CMB, this much shorter second phase of vacuum domination may serve to dilute unwanted fields (e.g. the moduli problem) as well as give rise to a second period of reheating of the universe (see e.g.~\cite{Lyth:1995hj,Lyth:1995ka}). A prime example of a second vacuum domination consists in a strongly supercooled first order phase transition (see e.g.~\cite{Barreiro:1996dx}). The latter often occurs in connection with the breaking of gauge symmetries. While supercooling does not arise for the electroweak phase transition, simple and well-motivated gauge extensions of the Standard Model can trigger a supercooled phase transition (see e.g.~\cite{Jaeckel:2016jlh,Jinno:2016knw,Addazi:2017gpt,Hashino:2018zsi,Croon:2018erz,Marzo:2018nov,Breitbach:2018ddu,Baratella:2018pxi,Azatov:2019png,Lewicki:2020azd}). In the regime of strong supercooling the latter reheats the universe a second time and releases great amounts of entropy which dilute the preexisting plasma. In the language of Eq.~\eqref{eq:alpha}, any model with $\alpha\gg 1$ reheats the universe when the vacuum energy is converted to radiation. Subsequently, there may be some residual radiation from before the phase transition, but most of the radiation content of the universe arises as a result of the reheating from the supercooled transition.  The Hot Big Bang in this case is associated with the supercooled phase transition rather than with the end of inflation.  

The reheating temperature $T_*$ after the supercooled phase must be high enough for BBN to take place, i.e., we again require $T_*>1.8\:\text{MeV}$~\cite{Hannestad:2004px,Hasegawa:2019jsa}. 
In addition, there arises a CMB constraint that the second vacuum domination should last $\lesssim 10$ e-folds. This constraint ensures that the scales observable in the CMB exited the horizon during standard inflation, and not during the second vacuum domination (which would be a problem since the perturbations generated during the second vacuum domination have a very different spectrum compared to what is observed in the CMB, see e.g.~\cite{Lewicki:2021xku}). In the specific example we consider below, the second vacuum domination lasts only $\sim 1$~e-fold such that the CMB constraint is easily satisfied.\footnote{Because of the additional e-folds due to the second period of vacuum domination, the production of perturbations on CMB observable scales occurs at a later point in inflation, farther down the inflaton potential, compared to the standard inflationary scenario (where there is no second vacuum-dominated epoch).  However, in the example considered below, this shift is very small.}

As a simple example we consider a U(1)-gauge extension of the Standard Model commonly referred to as the Abelian Higgs model. The Lagrangian containing the complex charged scalar field $\Phi$ and the U(1) vector field $A_\mu$ -- the dark photon -- reads\footnote{The Abelian Higgs model without an explicit mass term has also been considered in the context of the NANOGrav signal~\cite{Lewicki:2021xku}},
\begin{equation}\label{eq:abelianhiggs}
\mathcal{L} = - \frac{1}{4} F_{\mu\nu} F^{\mu\nu} + \left|D_\mu \Phi  \right|^2 - V(\Phi)\,,\qquad
V(\Phi) = - \mu^2 | \Phi|^2 + \lambda| \Phi|^4 + V_0\,,
\end{equation}
with $V_0 = \mu^4/(4\lambda)$. Here we employed the standard definitions of the field tensor $F^{\mu\nu}=\partial^\mu A^\nu - \partial^\nu A^\mu$ and the gauge covariant derivative $D_\mu = \partial^\mu - i g A^\mu$ with $g$ denoting the gauge coupling. For convenience we can express the complex scalar field in terms of the real scalar $\phi=\sqrt{2|\Phi|}$ (and a phase field).

In addition to the Lagrangian terms in Eq.~\eqref{eq:abelianhiggs} we invoke a (small) coupling between the Abelian Higgs sector and the Standard Model (e.g.\ through the Higgs and/or vector portal of the Standard Model). The latter establishes thermal equilibrium between both sectors in the early universe.

At zero temperature, the potential in Eq.~\eqref{eq:abelianhiggs} features a minimum with vanishing vacuum energy at $\phi = \mu/\sqrt{\lambda} \equiv v$. In this minimum the gauge symmetry is broken and the dark photon and the scalar receive masses of $m_A = g v$ and $m_\phi= \sqrt{2} \mu$ respectively. However, in the hot early universe, the induced thermal potential $\Delta V_{\text{thermal}}$ stabilizes the scalar field at $\phi=0$ and thereby restores the gauge symmetry. As the universe cools down and temperature effects decrease, $\phi$ either rolls or tunnels into its symmetry-breaking minimum in a crossover or a phase transition. Today, the universe resides in the broken phase. 

Considering the full thermal potential of the Abelian Higgs model~\cite{Dolan:1973qd,Dine:1992wr,Arnold:1992rz} it turns out that a supercooled first order phase transition arises if the transition temperature $T_n$ is small compared to the dark photon mass in the true vacuum $T_n \ll m_A$~\cite{Niedermann:2021ijp,Niedermann:2021vgd}. As shown in these references as well as illustrated below (see the discussion following Eq.(\ref{eq:rat})), this situation is realized if $\lambda \ll g^4$ -- a relatively mild constraint given $g$ is of order unity. In the low-temperature/ high-mass regime the thermal potential can be written as~\cite{Niedermann:2021vgd}
\begin{equation}
\Delta V_{\text{thermal}}(\phi)\simeq 3 T^4 K(g\phi/T) e^{-g\phi/T}\,,
\end{equation}
where we skipped field-independent terms. The function $K$ is approximated by the following fit~\cite{Niedermann:2021vgd},
\begin{equation}
K(x)\simeq -0.1134 (1+x) - 0.113 x^2 + 4.32\times 10^{-6} \log(x) x^{3.58} + 0.0038 e^{-x(x-1)}\,.
\end{equation}

\begin{figure}[htp]
\begin{center}
\includegraphics[width=0.45\textwidth]{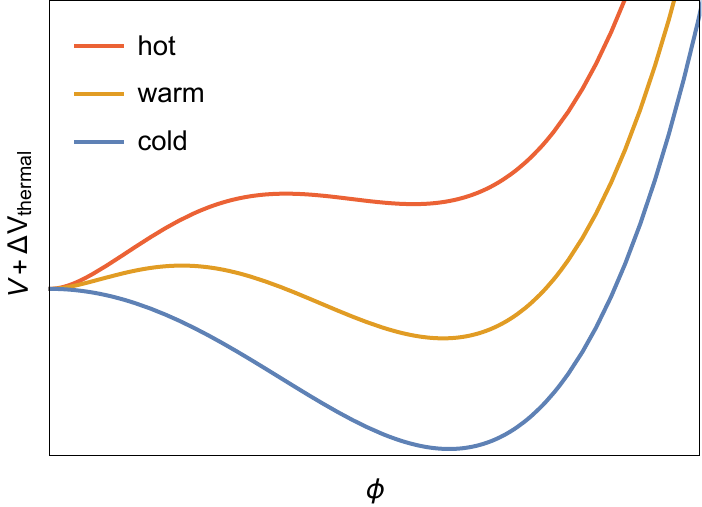}
\end{center}
\vspace{-5mm}
\caption{Finite temperature potential in the Abelian Higgs model. At high temperature (red curve) the universe settles in the gauge-symmetry preserving minimum at $\phi=0$. As the universe cools down, the symmetry-breaking minimum at $\phi=0$ shows up and eventually gets energetically preferred (orange curve). Once the barrier between the two minima has sufficiently decreased $\phi$ tunnels into the deeper minimum triggering a first order phase transition. At low temperature, $\phi=0$ becomes a maximum in the potential (blue curve).}
\label{fig:vthermal}
\end{figure}

In Fig.~\ref{fig:vthermal} we depict the full potential $V_{\text{tot}}= V(\Phi)+ \Delta V_{\text{thermal}}(\phi)$ including the zero-temperature and thermal parts at different temperatures. In the hot early universe the global minimum is located at $\phi=0$. But as the universe cools down the minimum at $\phi\neq 0$ shows up and eventually becomes energetically preferred. The thermal transition rate from the symmetry-preserving into the symmetry-breaking minimum is given by (cf. Eq.~\eqref{eq:tunnelingrate})
\begin{equation}
\Gamma = T^4 \left(\frac{S_3}{2\pi\,T}\right)^{3/2} e^{-S_3/T}\,.
\end{equation}
The Euclidean action $S_3$ needs to be determined numerically by solving the differential equation of the bounce. For simplicity we consider the case $\lambda\ll 1$ for which $S_3$ becomes independent of $\lambda$. In this regime, we find that the following fit function agrees well with the full numerical result,
\begin{equation}\label{eq:S3Tfit}
\frac{S_3}{T} \simeq \frac{1}{g^3}\left[ 603.4 \left(\frac{g\,T}{2\mu}-1\right)^{1.8}+344.3 \left(\frac{g\,T}{2\mu}-1\right)^3\right]\,.
\end{equation}

The (inverse) duration of the phase transition is obtained from Eq.~\eqref{eq:beta2},
\begin{equation}
\label{eq:dur}
 \frac{\beta}{H_*} \simeq  \left.\frac{d(S_3/T)}{H_*\,dt}\right|_{t=t_*} = \left.\frac{d(S_3/T)}{H_*\,dT}\dot{T}\right|_{T=T_n} \simeq T_n \left.\frac{d(S_3/T)}{dT}\right|_{T=T_n}\,.
\end{equation}
Here we used that the time-dependence of $\Gamma$ dominantly arises through the temperature-dependence of $S_3/T$. 
The radiation temperature $T_n$ right before the phase transition is fixed by Eq.~\eqref{eq:Gammatstar}. Combining Eqs.(\ref{eq:S3Tfit}) and (\ref{eq:dur}), we see  that $\beta/H_*$ decreases monotonically for growing $g$. Imposing a perturbative gauge coupling strength $g^2<4\pi$, therefore, leads to the constraint
\begin{equation}\label{eq:betaHmin}
 \frac{\beta}{H_*} \gtrsim 500\,.
\end{equation}
As a reminder, the minimal value of $\beta/H_*$ (here corresponding to the maximal value of gauge coupling strength) corresponds to the largest gravitational wave amplitude, see Eq.~\eqref{eq:gravityspectrumplasma}.
Once $T_n$ is known, the ratio of vacuum-to-radiation energy follows from Eq.~\eqref{eq:alpha},
\begin{equation}
\label{eq:rat}
\alpha = \frac{V_0}{\rho_{\text{r}}(T_n)} = \frac{\mu^4/(4\lambda)}{(\pi^2/30)g_{\text{eff}}(T_n)\,T_n^4}\,.
\end{equation}
Independent of the coupling choice we find that $T_n=(2-10)\times(\mu/g)$ which implies $\alpha=\mathcal{O}(g^4/\lambda)$. This confirms that the regime of strong supercooling ($\alpha\gg 1$) is indeed accessed for $\lambda \ll g^4$.

In Tab.~\ref{tab:scparameters} we provide an example parameter choice yielding $\alpha = 14.5$. For this large value of $\alpha$, the universe was strongly vacuum-dominated just before the phase transition: $V_0$ made up 94\% of the energy density of the universe and $\rho_r$ made up 6\%. 
The phase transition then converts $V_0$ to a new (dominant) component of radiation.
Thus most of the radiation density of the present universe is produced by the supercooled phase transition which hence plays the role of the Hot Big Bang.
\footnote{We note the actual phase transition
was virtually instantaneous, with duration $H_*/\beta = 1/720 = 0.014$ e-folds.} 

The number of e-foldings of the scale factor during the vacuum-dominated epoch is approximately given by ${\rm log} (T_*/T_n)$. If the transition time is short (compared to the Hubble time) we can approximate Eq.~\eqref{eq:Tst} as
$T_* \simeq (1+\alpha)^{1/4}\,T_n$.
For the choice $\alpha=14.5$ 
the epoch of vacuum domination (during the supercooling stage) 
produces roughly one e-fold of 
expansion.

The addition to the Lagrangian of even a small portal coupling (of the dark sector to the Standard Model) will be sufficient to ensure that -- in the symmetry breaking vacuum -- the Abelian Higgs fields decay promptly into electromagnetic radiation (given the decay to electron-positron pairs is kinematically accessible). Therefore, we can assume that the phase transition also reheats the visible sector which subsequently evolves according to the cosmological standard model.\footnote{The condition $m_\phi > 2m_e$ also ensures that $\phi$ does not (significantly) alter the number of relativistic species during BBN.} Baryons and dark matter may either be produced in the phase transition, or may stem from the preexisting plasma.\footnote{Baryons and dark matter present in the preexisting plasma also get diluted by the phase transition. However, in their case, the entropy production can be compensated by enhancing the baryon/ dark matter fraction prior to the phase transition.}   

In the supercooled regime the true-vacuum bubbles propagate (virtually) at the speed of light. However, since the phase transitions involves the breaking of a gauge symmetry, the bubble walls experience a pressure which grows linearly with their Lorentz boost~\cite{Bodeker:2017cim}. Unless the supercooling is extremely strong (which would require $\alpha\gg 10^5$) the bubble walls do not reach the runaway regime in which they carry most of the energy density upon collision. Instead most of the available energy gets converted into plasma bulk motion and thermal energy. Hence, the gravitational wave signal from bubble collisions is suppressed.  On the other hand, the interactions of the bubble walls with the plasma induce sound waves which themselves source gravitational waves. The corresponding spectrum is determined by Eq.~\eqref{eq:gravityspectrumplasma} with $\kappa_v\simeq 1$ for $\alpha\gg 1$~\cite{Espinosa:2010hh}. Notice that, in contrast to the gravitational waves from bubble collisions, the peak amplitude is only suppressed by one power of $H_*/\beta$. Therefore, the range of $\beta$ consistent with the pulsar timing signals is slightly extended in the case of acoustic gravitational waves.

\begin{table}[htp]
\begin{center}
\begin{tabular}{|ll|ll|}
\hline
&&&\\[-4mm]
\multicolumn{2}{|c|}{Input Parameters}& \multicolumn{2}{c|}{Phase Transition} \\
 \hline
&&&\\[-4mm] 
$g$~[GeV] & $2$ & $T_*$~[MeV] & $5.0$\\[1mm]
$\lambda$ & $2\times 10^{-4}$ & $\beta/H_*$ & $720$\\[1mm]
$\mu$~[MeV] & $1.2$ & $\alpha$ & $14.5$\\
\hline
\end{tabular}
\end{center}
\vspace{-0.4cm}
\caption{Example choice of couplings and mass in the Abelian Higgs model (cf.~Eq.~\eqref{eq:abelianhiggs}) resulting in a supercooled phase transition. The phase transition parameters are also given. See Fig.~\ref{fig:spectra2} (line P5) for the corresponding gravitational wave spectrum.}
\label{tab:scparameters}
\end{table}

In Fig.~\ref{fig:spectra2} (line P5) we depict the acoustic gravitational wave spectrum of the Abelian Higgs model with the parameter choice of Tab.~\ref{tab:scparameters}. We have chosen a large value of the gauge coupling in order to minimize the suppression of the peak amplitude by $ H_*/ \beta$ (see discussion around Eq.~\eqref{eq:betaHmin}). The obtained spectrum falls in the right range to explain the pulsar timing signals with the normalization a bit low in the first NANOGrav bin (and similar for the other pulsar timing arrays). We note, however, that the fit can potentially be further improved by including the gravitational waves from magneto-hydrodynamic turbulence induced by bubble collisions. While the magnitude of this contribution is somewhat uncertain it is expected to soften the infrared tail of the spectrum which is favorable for fitting the NANOGrav, PPTA and EPTA signals. We also emphasize that further parameter space can be accessed in gauge extensions beyond the Abelian Higgs model. In this light a supercooled Big Bang phase transitions provides an attractive explanation for the pulsar timing signals.

\subsection{A Dark Big Bang}\label{sec:darkbigbang}

We have so far described a number of cosmological scenarios featuring a Hot Big Bang phase transition consistent with the NANOGrav, PPTA and EPTA signals. In this section we will turn to a complementary case, in which visible radiation/ matter and dark matter are of different origin. While the Hot Big Bang at the end of inflation generates the Standard Model plasma, dark matter is only produced much later in a `Dark Big Bang' -- a first order phase transition in the dark sector. In the following we will consider the Dark Big Bang (rather than the Hot Big Bang) as the explanation for the signals observed by the pulsar timing array experiments. Related ideas of linking dark phase transitions, dark radiation and pulsar timing signals have appeared in~\cite{Schwaller:2015tja,Nakai:2020oit,Addazi:2020zcj,Ratzinger:2020koh,Borah:2021ocu,Lewicki:2021xku}.
Below, we will assume that inflation and reheating to the visible sector has already taken place at an earlier epoch in the Universe, prior to the Dark Big Bang phase transition described here.

In a minimal realization, the dark sector is comprised of the tunneling field $\varphi$, the dark matter field $\psi$ and one/ several massless (or very light) degrees of freedom $\xi_i$ playing the role of dark radiation. The particle nature of $\psi$ is irrelevant for the following discussion, but for concreteness we will take $\psi$ to be a Majorana fermion. Furthermore, we assume that the dark sector is decoupled from ordinary matter (other than through gravity). The dark sector Lagrangian reads,
\begin{align}\label{eq:LDS}
\mathcal{L}_{\text{DS}}&= \frac{1}{2}\partial_\mu \varphi\partial^\mu \varphi  - V(\varphi)
+ \frac{i}{2}\bar{\psi}\cancel\partial\psi - \frac{m_\psi}{2} \bar{\psi}\psi- \kappa\, \varphi \bar{\psi}{\psi}+ \mathcal{L}_{\text{DR}}\,,\nonumber\\
V(\varphi)&= \frac{m_\varphi^2}{2}\varphi^2 - \mu \varphi^3 + \lambda^2 \varphi^4 + V_0\,,
\end{align}
where $\mathcal{L}_{\text{DR}}$ contains kinetic and interaction terms  of the dark radiation fields (self-interactions as well as interactions with the other dark sector fields). We left this Lagrangian part unspecified since it merely enters the early universe dynamics by fixing the annihilation cross section $\langle \sigma v \rangle_\psi$ of dark matter into dark radiation. In the following, we will simply take $\langle \sigma v \rangle_\psi$ to be a free parameter. The potential exhibits a false vacuum with energy density $V_0$ at $\varphi=0$ and the true vacuum at $\varphi=(3\mu+\sqrt{9\mu^2-16\lambda^2 m_\chi^2})/(8\lambda^2)$.\footnote{We assumed $\mu > 4\lambda m_\varphi/3$.} We chose $V_0$ such that the potential energy vanishes in the true minimum. 

Let us now turn to the cosmological evolution. We assume a standard inflationary epoch followed by the Hot Big Bang. The latter creates a thermal plasma of Standard Model particles, while reheating to dark sector particles is taken to be absent (or suppressed).\footnote{This is a natural choice since comparable reheating of both sectors would require a very non-generic choice of inflaton couplings.} Due to the absence of couplings to visible matter the dark sector remains cold for some time. The universe is assumed to populate the metastable minimum of $\varphi$ after inflation.\footnote{This situation is realized if inflation blows up a false vacuum patch to contain the entire observable universe}. The false vacuum energy is negligible at the beginning of the radiation-dominated epoch, but becomes more significant with time due to its slower redshift.

Long after the Hot Big Bang, at the time $t_*$, $\varphi$ tunnels into the true vacuum in a first order phase transition. We call this instant the `Dark Big Bang' since it creates a hot plasma of dark sector fields. Henceforth subscript $*$ refers to the time right after the Dark Big Bang phase transition, $T$ refers to the temperature of the visible sector, and $T_d$ refers to the temperature of the dark sector. Since the phase transition is fast compared to the Hubble time (which we will show below) we can estimate the dark sector temperature $T_{d*}$  right after the Dark Big Bang by setting
\begin{equation}
\rho_{\text{vac}}= V_0 = \frac{\pi^2}{30}g_d(T_{d*})T_{d*}^4\,,
\end{equation}
where $g_d$ counts the relativistic dark sector degrees of freedom which include $\xi_i$, $\psi$ and possibly $\varphi$ (depending on its mass). At the same time, the phase transition does not cause any entropy transfer from the dark to the visible sector due to the absence of any direct couplings. Hence, the temperature $T$ of the Standard Model plasma is not affected by the Dark Big Bang (implying $T_*=T_n$), where $T_n$ was the temperature just before the phase transition.). 

Parametrizing the ratio of vacuum to visible-radiation density at the phase transition by $\alpha$ as in Eq.~\eqref{eq:alpha} we can relate $T_{d*}$ and $T_*$,
\begin{equation}\label{eq:TdsTs}
 \frac{T_{d*}}{T_*}=\alpha^{1/4}\left(\frac{g_{\text{eff}}(T_*)}{g_d(T_{d*})}\right)^{1/4}\,.
\end{equation}
During the subsequent evolution of the universe the entropies of visible and dark sector are separately conserved. Therefore, the temperature ratio remains approximately constant up to changes in the effective number of degrees of freedom,
\begin{equation}\label{eq:TdT}
 \frac{T_{d}}{T} = \left(\frac{g_{\text{eff}}(T)}{g_{\text{eff}}(T_*)}\right)^{1/3}\left(\frac{g_d(T_{d*})}{g_d(T_d)}\right)^{1/3}\frac{T_{d*}}{T_*}\,.
\end{equation}
It is convenient to express the dark radiation density as an extra contribution to the effective neutrino number. Employing~Eq.~\eqref{eq:TdsTs} and~\eqref{eq:TdT} one finds (see also~\cite{Nakai:2020oit}),
\begin{equation}
\Delta N_{\text{eff}} = 0.63\times \left(\frac{\alpha}{0.1} \right)\left(\frac{10}{g_{\text{eff}}(T_*)} \right)^{1/3}\left(\frac{g_{d}(T_{d*})}{g_{d}(T_{d})} \right)^{1/3}\,.
\end{equation}
Planck data combined with local measurements of the Hubble constant suggest $\Delta N_{\text{eff}}=0.22\pm 0.15$. While a small dark radiation contribution to $N_{\text{eff}}$ is allowed (and even marginally preferred), the latter should not exceed $\Delta N_{\text{eff}} = 0.5$. For a phase transition at the MeV-GeV scale (i.e.\ in the frequency band of pulsar timing arrays) we, therefore, obtain the constraint
\begin{equation}\label{eq:alphamax}
\alpha \lesssim 0.1\,.
\end{equation}
We can conclude that the universe needs to be radiation-dominated at the time of the Dark Big Bang.

In order to determine the gravitational wave signal from the Dark Big Bang we need to express the phase transition parameters $\alpha$, $T_*$ and $\beta$ in terms of the Lagrangian parameters in Eq.~\eqref{eq:LDS}. Since we are considering a quartic potential of the tunneling field, we can use Eq.~\eqref{eq:tunneling_doublefield} (with $m_{\chi,\text{eff}}$ replaced by $m_\varphi$) to derive the tunneling rate $\Gamma$. The latter then fixes the time of the phase transition by the condition $I(t_*)=1$ with the integral $I$ as defined in Eq.~\eqref{eq:prob}. We can pull $\Gamma$ out of the integral since it has no time dependence. In evaluating the integral,  we can approximate the time-dependence of the scale factor by $a\propto t^{1/2}$, since the Dark Big Bang occurs during radiation domination (cf.\ Eq.~\eqref{eq:alphamax}). Thus the condition $I(t_*)=1$ implies
\begin{equation}\label{eq:tstar}
t_*\simeq\left(\frac{105}{8\pi\,\Gamma}\right)^{1/4}\,.
\end{equation}
Employing the time-temperature relation of radiation-domination we, furthermore, obtain
\begin{equation}
T_* \simeq \left(\frac{45\,M_{\text{P}}^2}{2\pi^2\,g_{\text{eff}}(T_*)\,t_*^2}\right)^{1/4}\,,
\end{equation}
and
\begin{equation}
\alpha\simeq \frac{4 \,t_*^2 V_0}{3 M_{\text{P}}^2}\,.
\end{equation}
For a given $\alpha$, the duration of the phase transition is obtained from Fig.~\ref{fig:betaalpha}. 

During the Dark Big Bang phase transition bubbles of true vacuum are formed. Since the dark sector is decoupled from the Standard Model plasma, the expansion of the bubbles is not affected by the surrounding plasma. Therefore, the bubble walls can reach the runaway regime in which the entire gravitational wave signal stems from the bubble collisions (while acoustic gravitational waves are absent). The gravitational wave spectrum from the Dark Big Bang is thus determined by Eq.~\eqref{eq:gravityspectrum} with $\kappa_\phi=1$.

The dark matter abundance in the Dark Big Bang scenario can be set by a thermal freeze-out  in the dark sector~\cite{Feng:2008mu}.
After the bubble walls have collided, the dark sector quickly reaches a thermal state with temperature $T_{d*}$ given by~Eq.~\eqref{eq:TdsTs}.\footnote{The evolution of a universe with decoupled visible and dark sectors at different temperatures has been studied in the context of asymmetric reheating~\cite{Hodges:1993yb,Berezhiani:1995am,Adshead:2016xxj}.}
The dark plasma contains the dark radiation degrees of freedom $\xi_i$ and the dark matter field $\psi$ (which we assume to be lighter than $T_{d*}$).\footnote{Quanta of the tunneling field $\phi$ may initially also be contained in the plasma, but decay away quickly to other dark sector particles. Since $m_{\phi}$ is typically of the same order as $T_{d*}$, the $\phi$ particles are nonrelativistic after the phase transition and their abundance is suppressed.} Reactions $\psi\psi \leftrightarrow \xi_i\xi_i$ keep dark matter in thermal equilibrium (approximately) until the Hubble rate of expansion drops below the dark matter annihilation rate. At this moment $\psi$ freezes out and the total number of $\psi$ particles remains fixed. We denote the freeze-out dark sector temperature by $T_{d,f}$.

It is convenient to introduce the abundance as the ratio of $\psi$ number density over dark entropy density, $\Upsilon_\psi= n_\psi/s_{\text{dark}}$. Employing dark entropy conservation, the Boltzmann equation for $\Upsilon_\psi$ takes the form~\cite{Lee:1977ua},
\begin{equation}\label{eq:boltzmann}
 \frac{d\Upsilon_\psi}{dx}=-\frac{(\sigma v)_\psi \,s_{\text{dark}}}{Hx}\left(\Upsilon_\psi^2-\Upsilon_{\psi,eq}^2\right)\,,
\end{equation}
where we introduced $x=m_\psi/T_d$. Notice that the only way the visible sector enters Eq.~\eqref{eq:boltzmann} is by contributing to the Hubble expansion rate.

The equilibrium abundance $\Upsilon_{\psi,eq}$ can be obtained by integrating the Fermi-Dirac distribution. In the following we focus on a freeze-out in the non-relativistic regime ($x_{f}=m_\psi/T_{d,f}\gtrsim 3$) which allows us to approximate
\begin{equation}
\Upsilon_{\psi,eq} = \frac{45}{4\pi^4} \frac{g_\psi\, x^2\,K_2(x)}{g_d(x)}\,,
\end{equation}
where $g_\psi$ counts the internal degrees of freedom ($g_\psi=2$ for a Majorana fermion) and $K_2$ stands for the second modified Bessel function of the second kind.

The solution to Eq.~\eqref{eq:boltzmann} initially follows the equilibrium abundance before smoothly turning into a constant at the time of freeze-out. The terminal abundance $\Upsilon_{\psi}(\infty)$ can be found by solving Eq.~\eqref{eq:boltzmann} numerically. The corresponding relic density of $\psi$-particles reads
\begin{equation}\label{eq:omegah2dm}
\Omega_{\psi} h^2= \frac{m_\psi \,\Upsilon_{\psi}(\infty)\,s_{\text{dark}}(T_{d,0})}{3 (H_0/h)^2 M_{\text{P}}^2} = 2.74 \times 10^5 \:\left( \frac{m_\psi}{\text{MeV}}   \right)\,\alpha^{3/4} \left(\frac{g_d(T_{d*})}{g_{\text{eff}}(T_*)}\right)^{1/4}\: \Upsilon_{\psi}(\infty)\,,
\end{equation}
where $H_0/h=100\:\text{km}/(\text{s}\text{Mpc})$. In the second step, we employed Eq.~\eqref{eq:TdT} and today's visible sector temperature $T=2.73\:\text{K}$ to obtain the dark entropy. In a viable dark sector freeze-out scenario $\Omega_{\psi} h^2$ needs to match the observed dark matter relic density $\Omega_{\text{DM}} h^2 = 0.1198 \pm 0.0012$~\cite{Planck:2018vyg}. This imposes a constraint on the dark matter annihilation cross section $\langle\sigma v\rangle_\psi$.\footnote{By $\langle \sigma v \rangle_\psi$ we denote the thermally averaged cross section at the time of freeze-out.} For sizeable $\alpha$ (say $\alpha \gtrsim 10^{-3}$) we find that the required cross section is of order $\langle \sigma v \rangle_\psi=\mathcal{O}(10^{-26}\text{cm}^3/s)$ -- similar as for a standard WIMP (i.e. visible sector freeze-out) scenario. This is not surprising since the dark sector temperature is not too different from the visible sector temperature in this case.

An important distinction, however, is that the dark freeze-out scenario can successfully be implemented with dark matter masses $m_\psi<\text{MeV}$. Such low masses imply that dark matter contributes to the number of relativistic species at the time of BBN. If $\psi$ was part of the visible sector it would add (at least) a full degree of freedom\footnote{A relativistic particle in equilibrium with the Standard Model plasma increases $g_{\text{eff}}(T)$ by the number of internal degrees of freedom (multiplied by $7/8$ in the case of a fermion).} which is in conflict with BBN constraints. However, as $\psi$ resides in a colder dark sector its contribution to the total energy density is reduced by $\alpha$. Hence, a relativistic $\psi$ (and additional relativistic dark radiation) at the time of BBN is viable as long as $\alpha$ is sufficiently small. Since CMB bounds already impose $\alpha\lesssim 0.1$ (cf.~Eq.~\eqref{eq:alphamax}) BBN does not provide an additional constraint.

The fact that dark matter in this model receives the correct adiabatic density perturbations required by CMB observations will be shown in a followup paper.  Clearly the usual production of DM perturbations does not take place during inflation since the DM does not yet exist. Instead, perturbations in the visible sector that are produced during inflation can later be transmitted gravitationally to the dark matter.

\begin{table}[htp]
\begin{center}
\begin{tabular}{|ll|ll|}
\hline
&&&\\[-4mm]
\multicolumn{2}{|c|}{Input Parameters}& \multicolumn{2}{c|}{Cosmology} \\
 \hline
&&&\\[-4mm] 
$m_\varphi$~[MeV] & $26.04$ & $\Omega_\psi h^2$ & $0.119$\\[1mm]
$\mu$~[MeV] & $40.72$ & $\Delta N_{\text{eff}}$ & $0.40$\\
\cline{3-4}
&&&\\[-4mm]
$\lambda$ & $1$ &  \multicolumn{2}{c|}{$\;\;$Phase Transition$\;\;$}\\
\cline{3-4}
&&&\\[-4mm]
$m_\psi$~[MeV] & $0.200$ & $T_*$~[MeV] & $20$\\[1mm]
$g_d(T_d*)$ & $6.75$ & $\beta/H_*$& $7.8$\\[1mm]
$\langle \sigma v \rangle_\psi$~[$cm^3/s$]$\;\;$ & $1.74\times 10^{-26}\;$ & $\alpha$ & $0.06$\\
\hline
\end{tabular}
\end{center}
\vspace{-0.4cm}
\caption{Parameter example in the Dark Big Bang scenario containing the tunneling field $\varphi$, the dark matter field $\psi$ and light dark radiation fields (the model is defined in Eq.~\eqref{eq:LDS}). The resulting predictions for the dark matter relic density, dark radiation density (expressed in terms of $\Delta N_{\text{eff}}$) and phase transition parameters are shown on the right side. The resulting gravitational wave spectrum is depicted in Fig.~\ref{fig:spectra2} (line P4).}
\label{tab:dbbparameters}
\end{table}

\begin{figure}[htp]
\begin{center}
\includegraphics[width=0.5\textwidth]{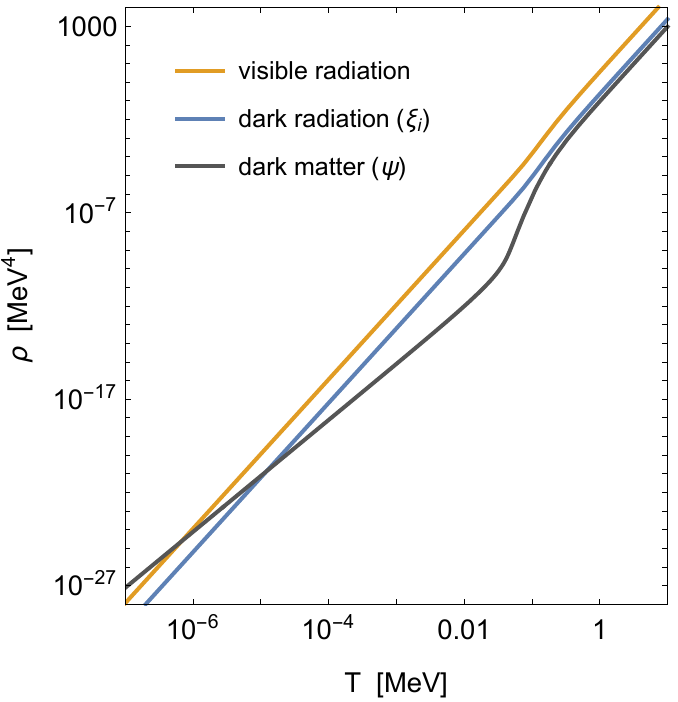}
\end{center}
\vspace{-5mm}
\caption{Evolution of the energy densities in visible radiation, dark radiation ($\xi_i$-particles) and dark matter ($\psi$-particles) after the Dark Big Bang phase transition. The parameters from Tab.~\ref{tab:dbbparameters} were assumed. The radiation densities (approximately) decrease as $T^4$. At high temperature $T\gtrsim 0.3\:\text{MeV}$ (corresponding to $T_d\gtrsim m_\psi$) the dark matter density evolves parallel to the radiation densities. But after dark matter becomes non-relativistic at $T\sim 0.3 \:\text{MeV}$ its density decreases exponentially until annihilation reactions freeze out at $T\sim 40 \:\text{keV}$. Below this temperature the number of dark matter particles remains fixed, implying that their energy density scales as $T^3$. At later times (beyond the left boundary of the plot), DM will eventually dominate over both types of radiation.}
\label{fig:darkfreeze}
\end{figure}

In Tab.~\ref{tab:dbbparameters} we provide a parameter example for the Dark Big Bang model defined in Eq.~\eqref{eq:LDS}. The example point features a Dark Big Bang phase transition at $T_*=20\:\text{MeV}$ which converts the dark vacuum energy into a hot dark plasma of $\xi_i$ and $\psi$ particles. In Fig.~\ref{fig:darkfreeze} we depict the evolution of the visible radiation, dark radiation ($\xi_i$) and dark matter ($\psi$) energy densities after the Dark Big Bang. Both radiation densities decrease as $T^4$ until the present epoch.\footnote{Slight deviations from $\rho_{\text{r}}\propto T^4$ occur due to changes in $g_{\text{eff}}(T)$.} The dark matter density $\rho_{\text{DM}}$ evolves parallel to the radiation densities as long as the $\psi$-particles are highly relativistic. But once $T_d \lesssim m_\psi$ the Boltzmann suppression sets in and $\rho_{\text{DM}}$ starts to decrease exponentially with $m_\psi/T_d$. Later, at $T_d\sim m_\psi/10$, annihilations become inefficient and the number of dark matter particles remains fixed. After the freeze-out $\rho_{\text{DM}}$ decreases with $T^3$ as in standard cold dark matter scenarios. For the example point, the relic density of $\psi$-particles agrees with the observed dark matter density. 
Apart from the dark plasma, the Dark Big Bang generates strong gravitational radiation. In Fig.~\ref{fig:spectra2} (line P4) we depict the gravitational wave spectrum for the parameter point in Tab.~\ref{tab:dbbparameters}. Since the benchmark point resides close to the thin-wall regime of vacuum tunnelling we expect the spectrum to follow approximately the prediction of the envelope approximation (left panel of the figure). As can be seen, a good fit to the NANOGrav signal is obtained. The benchmark point is also indicated in Fig.~\ref{fig:spectra} (P4 in the right panel), where one can see that it is also consistent with the PTA signal. Intriguingly, the Dark Big Bang explanation of the NANOGrav signal simultaneously predicts a non-negligible dark radiation density in the universe ($\Delta N_{\text{eff}}\sim 0.4$) which will be tested by future CMB experiments.

\section{Conclusion}\label{sec:conclusion}

The origin of the Hot Big Bang remains one of the big mysteries in cosmology. In this work we provided strong motivation that the Big Bang occurred through a strong first order phase transition. In this scenario the universe is initially trapped in a false vacuum which eventually decays through quantum tunneling. The latter triggers the formation of true vacuum bubbles in the sea of false vacuum. Bubble collisions generate a hot plasma of particles heralding the entrance into the radiation-dominated era. 

A common feature of all Big Bang first order phase transition cosmologies is the presence of strong gravitational radiation which is formed by the collision of true-vacuum bubbles. 
In this work we investigated, whether the Hot Big Bang could be responsible for the tentative observation of a stochastic gravitational wave background by the NANOGrav, PPTA and EPTA pulsar timing array experiments. By performing a fit to the pulsar timing array data we identified the range of phase transition temperatures, durations and strengths compatible with the signal (see Fig.~\ref{fig:spectra}). In particular, we found that the pulsar timing signals can be explained if the reheating temperature of the Hot Big Bang, and correspondingly the energy scale of the false vacuum, falls in the range $T_* \sim \rho_{{\rm vac}}^{1/4} =\text{MeV}-100\:\text{GeV}$.

The idea of a first-order Big Bang phase transition originally emerged within Guth's ``old inflation'' proposal. While the original model fails because of the empty-universe problem, modifications can reconcile the vacuum transition picture with cosmological data and --at the same time -- support the low vacuum-energy scale required to fit the pulsar timing signals. In Sec.~\ref{sec:scenarios} we present a number of well-motivated cosmologies with a successful Big Bang first order phase transition which reheats the universe -- either at the end of inflation, after a period of kination, or after a second period of vacuum-domination long after inflation:
\begin{itemize}
\item In Double Field Inflation (Sec.~\ref{sec:doublefield}) the tunneling field is coupled to a rolling field which catalyzes a very rapid first order phase transition (resolving the empty universe problem). We showed that the low inflation scale required to fit the pulsar timing signals can be accessed without running into the fine-tuning problems plaguing low-scale slow roll inflation. A low-scale double field version of $\alpha$-attractor inflation is introduced which satisfies all cosmological constraints.
\item Chain Inflation (Sec.~\ref{sec:chaininflation}) features a Universe tunneling through a series of ever lower vacuum energies. Each individual transition completes quickly within a fraction of a Hubble time (avoiding the empty universe problem), while all transitions together support sufficient e-foldings of inflation. Since the origin of CMB perturbations in chain inflation consists in the probabilistic nature of tunneling (rather than quantum fluctuations of the inflaton as in slow roll inflation) the low inflation scales favored by the pulsar timing arrays is shown to be accessed without the requirement of an extremely flat (i.e.\ tuned) potential (in contrast to slow roll inflation).
\item The proposed ``Kination-Induced Big Bang'' (Sec.~\ref{sec:kinBB}) corresponds to a strong first-order phase transition after a period of kinetic-energy domination of the universe. Such a kination period is predicted e.g.\ by quintessential inflation for which the Kination-Induced Big Bang provides a new reheating mechanism.
\item A ``Supercooled Big Bang'' (Sec.~\ref{sec:supercooledBB}) refers to a strongly supercooled thermal first-order phase transition. We present an example model in which the latter occurs after a short second period of vacuum-domination long after inflation and reheats the universe a second time.
\item Finally in Sec.~\ref{sec:darkbigbang}, we proposed that the Hot Big Bang at the end of inflation generates the Standard Model plasma, but dark matter is only produced much later in a ``Dark Big Bang'' – a first order phase transition in the dark sector. 
\end{itemize}
For the five complementary models with a Big Bang phase transition we derived the spectrum of gravitational waves and compared them to the pulsar timing signal (see Fig.~\ref{fig:spectra2}). In all cases we found parameter examples featuring a gravitational wave signal in agreement with the pulsar timing arrays. We concluded that a Big Bang phase transition provides an attractive explanation for the NANOGrav, PPTA and EPTA results.

Nevertheless, there is still a long way to establish the detection of a Big Bang first order phase transition. First, the unambiguous discovery of a stochastic gravitational wave background by NANOGrav, PPTA, EPTA or any other pulsar timing array experiment requires the measurement of the quadrupolar spatial correlations predicted by General Relativity. In the optimistic case -- since pulsar timing arrays are continuously improving their statistics -- the detection of the quadrupolar correlations could be just around the corner. If a gravitational wave signal is confirmed the Big Bang origin must be discriminated against other astrophysical and cosmological gravitational wave sources. In this respect it will be crucial to further improve the prediction of the gravitational wave spectrum from phase transitions beyond the simplified assumptions entering e.g.\ the envelope approximation. Moreover, it will be important to investigate complementary cosmological probes of a Big Bang phase transition. Such probes could include an increased $\Delta N_{\text{eff}}$ (see Sec.~\ref{sec:darkbigbang}), correlations between inflationary and phase transition observables -- the chain inflation scenario of Sec.~\ref{sec:chaininflation} e.g.\ correlates $n_s$ and $T_*$ -- as well as other impacts on BBN and CMB observables. 

The search for gravitational wave signals from a first order phase transition is not limited to pulsar timing arrays (see~\cite{Caprini:2019egz} for a recent review). With future space- and ground-based interferometers there is hope to detect a stochastic gravitational wave background in the $\text{mHz} - \text{kHz}$-regime. Simple estimates based on~Eq.~\eqref{eq:vacuumomega} and~\eqref{eq:vacuumf} suggest that (e)LISA can potentially probe a Big Bang first order phase transition with an energy density of the false vacuum $\rho_{\text{vac}}^{1/4}\sim (10^2-10^5)\:\text{GeV}$, while the next stage of LIGO-Virgo-KAGRA~\cite{Harry:2010zz,VIRGO:2014yos,Somiya:2011np} (or possibly next-generation experiments like Einstein Telescope~\cite{Punturo:2010zz} and Cosmic Explorer~\cite{Reitze:2019iox}) could access $\rho_{\text{vac}}^{1/4}\sim (10^8-10^9)\:\text{GeV}$. All the first order phase transition models presented in this paper can also produce gravitational waves detectable in these upcoming searches.

Our findings motivate a dedicated experimental and theoretical program to test a Big Bang first order phase transition through the associated gravitational radiation. Needless to say that the prospect of directly probing the Hot Big Bang through its gravitational wave signature is extremely exciting.

\section*{Acknowledgements}
K.F.\ is Jeff \& Gail Kodosky Endowed Chair in Physics at the
University of Texas at Austin, and K.F.\ and M.W.\ are grateful for
support via this Chair. K.F.\ and M.W.\ acknowledge support by
the Swedish Research Council (Contract No. 638-2013-8993). 
This material is based upon work supported by the U.S. Department of Energy, Office of Science, Office of High Energy Physics program under Award Number DE-SC-0002424.

\bibliography{nanobib}
\bibliographystyle{h-physrev}

\end{document}